# Superconducting dome associated with the suppression and re-emergence of charge density wave states upon sulfur substitution in $CuIr_2Te_4$ chalcogenides

Mebrouka Boubeche [a], Ningning Wang[b], Jianping Sun [b], Pengtao Yang [b], Lingyong Zeng [a], Shaojuan Luo [c], Yiyi He [a], Jia Yu [d], Meng Wang [d], Jinguang Cheng [b], Huixia Luo [a*]

[a] School of Materials Science and Engineering, State Key Laboratory of Optoelectronic Materials and Technologies, Key Lab of Polymer Composite & Functional Materials, Guangzhou Key Laboratory of Flexible Electronic Materials and Wearable Devices, Sun Yat-Sen University, No. 135, Xingang Xi Road, Guangzhou, 510275, P. R. China

[b] Beijing National Laboratory for Condensed Matter Physics and Institute of Physics, Chinese Academy of Sciences and School of Physical Sciences, University of Chinese Academy of Sciences, Beijing 100190, China.

[c] School of Chemical Engineering and Light Industry, Guangdong University of Technology, Guangzhou, 510006, P. R. China

[d] Center for Neutron Science and Technology, School of Physics, Sun Yat-Sen University, Guangzhou, 510275, China

[*] Author to whom any correspondence should be addressed.

E-mail: luohx7@mail.sysu.edu.cn

## Abstract

We report the path from the charge density wave (CDW)-bearing superconductor $CuIr_2Te_4$ to the metal insulator transition (MIT)-bearing compound $CuIr_2S_4$ by chemical alloying with the gradual substitution of S for Te. The evolution of structural and physical properties of the $CuIr_2Te_{4-x}S_x$ ($0 \leq x \leq 4$) polycrystalline system is systemically examined. The X-ray diffraction (XRD) results imply $CuIr_2Te_{4-x}S_x$ ($0 \leq x \leq 0.5$) crystallizes in a NiAs defected trigonal structure, whereas it adapts to the cubic spinel structure for $3.6 \leq x \leq 4$ and it is a mixed phase in the doping range of $0.5 < x < 3.6$. Unexpectedly, the resistivity and magnetization measurements reveal that small-concentration S substitution for Te can suppress the CDW transition, but it reappears around $x = 0.2$, and the CDW transition temperature enhances clearly as $x$ augments for $0.2 \leq x \leq 0.5$. Besides, the superconducting critical temperature ($T_c$) first increases with S doping content and then decreases after reaching a maximum $T_c = 2.82$ K for $CuIr_2Te_{3.85}S_{0.15}$. MIT order has been observed in the spinel region ($3.6 \leq x \leq 4$) associated with $T_{MI}$ increasing with $x$ increasing. Finally, the rich electronic phase diagram of temperature versus $x$ for this $CuIr_2Te_{4-x}S_x$ system is assembled, where the superconducting dome is associated with the suppression and re-emergence of CDW as well as MIT states at the end upon sulfur substitution in the $CuIr_2Te_{4-x}S_x$ chalcogenides.

**Introduction:**

The family of ternary chalcogenides has been studied intensively due to their rich structural and physical properties. [1-3] Notably, the copper chalcogenides with cubic spinel structure exhibit rich quantum states, including metal-insulator transition (MIT), magnetism, superconductivity (SC) and so on. [4-8] Normal thiospinel $CuIr_2S_4$ is of particular interest because it sustains a MIT at around 230 K under normal pressure, where the energy gap for the insulating phase is around 0.094 eV. [8] So far, many researchs on the effect of chemical doping or adding physical pressures on the MIT in $CuIr_2S_4$ have been reported. In most cases, external disturbances include chemical doping (such as $CuIr_2S_{4-x}Se_x$, $CuIr_{2-x}Ti_xS_4$, $CuIr_{2-x}Rh_{x2}S_4$, $CuIr_{2-x}Pt_xS_4$, $Cu_{1-x}Ag_xIr_2S_4$, $Cu_{1-x}Ni_xIr_2S_4$, $Cu_{1-x}Zn_xIr_2S_4$) and adding physical pressures disrupt the MIT and even further induce SC in $CuIr_2S_4$. [9-18]

On the other hand, $CuIr_2Te_4$ adopting a NiAs defected structure with a trigonal symmetry features the occurrence of SC and charge density wave (CDW)-like transition, characterized by magnetization and resistivity measurements. [19,20] The first-principles calculation analysis implies that the density of states (DOS) near the Fermi energy for $CuIr_2Te_4$ mainly originates from the Ir *d* and Te *p* orbitals. [20] Subsequently, it is experimentally proved that the CDW and SC are both sensitive to the diverse chemical dopants and the three doping sites (Cu-site, Ir-site and Te-site) in the pristine $CuIr_2Te_4$. [22-24] Therefore, it is still of interest to explore the path from the layer $CuIr_2Te_4$ chalcogenide to the $CuIr_2S_4$ spinel by chemical alloying.

Here, we focus on studying the substitution of S for Te in the $CuIr_2Te_4$ host material based on the following aspects: (i) there is a big difference between the structures of two end compounds — $CuIr_2S_4$ crystallizes in a cubic spinel structure, but $CuIr_2Te_4$ adopts a layered structure. (ii) Two end compounds show distinct physical properties — $CuIr_2Te_4$ shows the coexistence of CDW-like transition and SC, whereas $CuIr_2S_4$ exhibits MIT. (iii) S belongs to the same chalcogen family as Te but has a smaller ionic radius. Therefore, these aforementioned significant differences between the two compounds engendered us to probe the path between $CuIr_2Te_4$ and $CuIr_2S_4$ by chemical alloying, in which rich structural and physical properties are expected in the new $CuIr_2Te_{4-x}S_x$ ($0 \leq x \leq 4$) solid solution. The structural and physical properties of $CuIr_2Te_{4-x}S_x$ ($0 \leq x \leq 4$) are characterized through XRD, resistivity, magnetization and heat capacity tests.

## Experimental Section

Polycrystalline specimens of $CuIr_2Te_{4-x}S_x$ ($0 \leq x \leq 4$) were synthesized from the stoichiometric admixture of high-purity elements of Cu (99.999%), Ir (99.99%), Te (99.99%) and S (99.9%) with 0.1 wt% excess of Te and S. The mixtures were sealed in evacuated quartz tubes and heated up to 850 $^{o}$C (1 $^{o}$C/min) for four days, followed by furnace cooling with the rate of 3 $^{o}$C/min to room temperature. Subsequently, the resultant specimens were ground and heated in pelletized cylinder form at 850 $^{o}$C (3 $^{o}$C/min) for another five days.

Powder X-ray diffraction (PXRD, MiniFlex, Rigaku) with Cu Kα1 radiation was used to examine the phase structure of the $CuIr_2Te_{4-x}S_x$ ($0 \leq x \leq 4$) compounds. Rietveld refinements using the FULLPROF suite software were carried out to acquire the lattice parameters. [25] The element ratios were confirmed by scanning electron microscope combined with energy-dispersive x-ray spectroscopy (SEM-EDXS, COXEM EM-30AX). Quantum design physical property measurement system (PPMS was used to measure the electrical resistivity (four-probe method) ($\rho(T)$) on rectangular samples (5 x 1 x 0.8 mm$^3$) down to 1.8 K and heat capacity measurement was performed in the range of 1.8 K - 15 K. Quantum interference device (SQUID) Quantum Design MPMS system was used to measure the magnetic susceptibilities ($\chi(T,H)$). $T_c s$ were extracted from the average of superconducting transition region in $\rho(T)$ data and the extrapolation point of the steep slope of the superconducting state and the normal state susceptibility, $T_c$ was also determined from the specific heat capacity $C_p$(T) from the equal area entropy construction. The values of $T_{MI}$ were determined from the inflection of the $\rho$(T) and $\chi$(T) curves at high temperatures.

## Results and Discussion

**Fig. 1(a-b)** and **Table 1** exhibit the refinement results of the XRD measurement for the selected layered $CuIr_2Te_{3.85}S_{0.15}$ and cubic $CuIr_2Te_{0.1}S_{3.9}$ compounds. The reflection peaks of $CuIr_2Te_{3.85}S_{0.15}$ are corresponding to the trigonal phase of $CuIr_2Te_{4-x}S_x$ having the space group *P*-3m1 (the inset of **Fig. 1(a)),** whereas the $CuIr_2Te_{0.1}S_{3.9}$ sample has a cubic spinel phase with the space group *Fd*-3m (the inset of **Fig. 1(b)**). Small amount of unreacted Ir exists in all studied specimens. Rietveld refinements for the other examined compounds are represented in **Fig. S1** in online supplemental information. From **Fig. 2(a-b)** and **Fig. S2,** it is evident that there are three distinct crystal structure zones at $0 \leq x \leq 0.5$, $0.5 < x < 3.6$ and $3.6 \leq x \leq 4$, respectively. It can be seen that the obtained powder diffractograms for $CuIr_2Te_{4-x}S_x$ ($0 \leq x \leq 0.5$) samples are mainly indexed to the trigonal phase. In zone $0.5 < x < 3.6$, the cubic spinel phase starts to appear, which indicates the coexistence of the layered and cubic phases (see **Fig. S2**). The main cubic spinel phase is obtained at the doping region of $3.6 \leq x \leq 4$.

In order to show the influence of sulfur substitution on the crystal structures, we magnified the (002) peaks for $CuIr_2Te_{4-x}S_x$ ($0 \leq x \leq 0.5$) and the (311) peaks for $CuIr_2Te_{4-x}S_x$ ($3.6 \leq x \leq 4$)) as depicted in **Fig. 2(a-b)**. We distinguished an obvious shift toward higher angles upon increasing the doping concentration *x* in the $CuIr_2Te_{4-x}S_x$ ($0 \leq x \leq 0.5$, $3.6 \leq x \leq 4$) compounds as presented in the inset of **Fig. 2(a-b)**, illustrating the incorporation of S into the $CuIr_2Te_4$ lattice. This shift is linked to the change on lattice constants *a* and *c*, as presented in **Fig. 2(c)**. Obviously, both lattice parameters *a* and *c* for the $CuIr_2Te_{4-x}S_x$ ($0 \leq x \leq 0.5$) decrease as the doping content *x* increases, adapting to Vegard's law [26]. The decreasing trend of the lattice parameters with increasing S content up to $x = 0.5$ indicates that the doped S atoms substitute for the Te atoms in the lattice. Besides, in the spinel structure region ($3.6 \leq x \leq 4$), the peaks also gradually shift to higher angles. From **Fig. 2(c)**, we can see that the lattice parameters ($a = b = c$) in the cubic phase are also reduced since the ionic radius of $Te^{2-}$ (2.21 Å) is larger than that of the $S^{2-}$ ion (1.8 Å) [27]. Further, SEM-EDXS is performed to explore the morphology and atomic ratio (see **Fig. S3 and Fig. S4**) for the light doping powder samples. From **Fig. S3**, we can see that all the elements are evenly distributed in the powder samples. In addition, the obtained atomic ratios (see the inset of **Fig. S4**) are very close to those in the target compositions. Now we turn to investigate the physical properties for $CuIr_2Te_{4-x}S_x$ ($0 \leq x \leq 0.5$) by combining the temperature-dependent resistivity $\rho(T)$ and magnetization $\chi(T)$ measurements. The temperature-dependent resistivity $\rho(T)$ is present in **Fig. S5**. The normalized resistivity $\rho/\rho_{300K}$ $(T)$ data are displayed in the main panel of **Fig. 3(a)**. The resistivity results suggest that tiny amounts of S substitution suppress the CDW transition ($T_{CDW}$) ($T_{\mathrm{CDW}}$ is defined by the minimum of $d\rho/d$T as shown in the inset of **Fig. 3(a)**).

$CuIr_2Te_{4-x}S_x$ ($0 \leq x \leq 0.5$) samples show metallic behaviors above 3 K. From **Fig. 3(b)**, abrupt superconducting transitions can be seen in the $\rho/\rho_{300K}$ curves of the $CuIr_2Te_{4-x}S_x$ ($0 \leq x \leq 0.4$) compounds at low temperatures. We have estimated the residual resistivity ratio (RRR = $R_{300K}/R_{5K}$) and the results are given in **Table 2**. RRR increases from 4.07 for the host sample to 5.28 for optimal samples ($x = 0.15$). Normally, there should be a decrease in value of RRR with doping as result of induced disorder. In this case, one possible way to understand this behavior of the increased RRR in the low region S-doping samples could be that the introduction of small amount of S might cause an improvement in the sublattice order. This would improve local chemical and electronic uniformity, resulting in the suppression of CDW in the low region S-doping samples. Similarly, this behavior has also been reported in Be doped FeSe superconducting system [28]. Meanwhile, S doped samples exhibit steep superconducting transitions in the range of $0 < x \leq 0.4$, indicating the homogeneity of the doped samples (see **Fig. 3(b)** and **Table 2**). Surprisingly, CDW transition feature is absent in $CuIr_2Te_{4-x}S_x$ ($0 < x < 0.2$), while it reappears for $0.2 \leq x \leq 0.5$ and the $T_{CDW}$ gradually raises with enhancing S doping concentration. Simultaneously the RRR gradually reduces from 5.28 for $x = 0.15$ to 1.8 for $x = 0.5$ **(see Table 2).** The reduction of RRR implies that higher S doping can significantly induce disorder and S ions are effective scattering centres, [29-31] which may possibly explain the recurrence of the CDW. SC was further investigated by the magnetization tests, from **Fig. 3(c)**, we can see the evolution of the $T_c$s, which is consistent with the resistivity data. The temperature-dependent normalized magnetic susceptibility $4\pi\chi(T)$ is getting smaller by increasing the S concentration $x$. However, no superconducting transition is detected in magnetic susceptibility data for $x$ = 0.5 down to 1.8 K, which agrees with the $\rho(T)$ data. **Fig. S7** shows the d$\chi$/dT vs T for the layer phase $CuIr_2Te_{4-x}S_x$ ($0.1 \leq x \leq 0.5$) samples. From **Fig. S7,** it can see the d$\chi$/dT transition is getting broader with the increasing of S doping content and the d$\chi$/dT transition vanishes when $x$ = 0.5. From resistivity and magnetic susceptibility results, we can see that only layered phases show SC with $T_c$ slightly ascending with $x$ increasing and attains the highest value of 2.82 K for $CuIr_2Te_{3.85}S_{0.15}$, which is rather higher than the $T_c$ obtained by the optimal Ru (2.79 K) and Al doping (2.75 K). [21,24] Subsequently, there is a small drop of $T_c$ with the increase of $x$ and a SC dome can be observed. To confirm the re-appearance of the CDW-like order for ($0.2 \leq x \leq 0.5$) samples, we further investigate the temperature dependent-magnetic susceptibilities $\chi(T)$ with heating and cooling under a 10 kOe magnetic field. The inset of **Fig. 3(d)** presents the cooling $d\chi/dT(T)$ revealing the $T_{CDW}$s, which are consistent with the resistivity data (see the inset of **Fig. 3(a),** as well as the curves obtained from the cooling process). It seems like that these investigated compounds with the layered structure near the two ends exhibit CDW-like transitions accompanied with the superconducting dome in the whole layered region, which is very similar to the electronic phase diagram of the recently reported $CuIr_2Te_{4-x}(I/Se)_x$ [23,32] and the 2H-$TaSe_{2-x}S_x$ ($0 \leq x \leq 2$) system, [33] where the disorder have a significant role in the tendency of SC and CDW. Alike phenomena have also been found in Tl-intercalated $Nb_3Te_4$ single crystals, [34] where the re-appearance of CDW is attributed to the chaos in the Nb chains. Moreover, $M_xTiSe_2$ ($M$ = Mn, Cr, Fe) doping series also show analogous phenomena — CDW state vanishes at the low intercalant concentrations is ascribed to the local deformations of Se–Ti–Se layers around introduced $M$ atoms and the re-emergence of the CDW state in the over-doped region is probably because of reducing of the deformation of Se-Ti-Se layers.[35] Similarly, it is reasonable to deduce that the CDW gap is possibly opened due to the disorder created with S element doping, leading to a reduced density of states (DOS) near the Fermi level ($E_F$) and the suppression of $T_c$.[36] But this foundation still requires further research indeed. After all, $T_c$*s* of the 2H-$TaSe_{2-x}S_x$ series are remarkably greater than those of the two end undoped 2H-$TaSe_2$ and 2H-$TaS_2$ compounds, but there is only subtle enhancement of $T_c$ at even the optimal doping composition $CuIr_2Te_{3.85}S_{0.15}$.

Next, we perform the temperature dependence of zero-field resistivity measurements with heating and cooling for the cubic spinel $CuIr_2Te_{4-x}S_x$ ($3.6 \leq x \leq 4$) to investigate the MIT, as exhibited in the main panel of **Fig. 4(a)**, it is clear that $T_{MI}$ raises with S concentration $x$ in the spinel phase as the inset of **Fig. 4(a)**. From **Fig. 4(a)**, it can be seen that the resisitivity slightly decreases with the decreasing temperature above MIT, but whereas it increases abruptly with the decreasig temperature below MIT. For example, it increases from around $1\times10^{-2}$ Ω·cm to $1\times10^{3}$ Ω·cm with cooling run for the composition $x = 3.9$. Below MIT, the resistivity displays insulating behavior, which is consistent with the previous report.[9] To further confirm the MIT, we performed the temperature-dependent magnetic susceptibility measurements with heating and cooling for the spinel compositions $CuIr_2Te_{4-x}S_x$ ($3.6 \leq x \leq 4$) with applied magnetic field 10 kOe. **Fig. 4(b)** presents the temperature-dependent magnetic susceptibility under H = 10 kOe for $CuIr_2Te_{4-x}S_x$ ($x$ = 3.6, 3.7, 3.8, 3.9 and 4) compositions. $\chi$ ($T$) behaviors for different compounds seems to be analogous, we can find that there is an obvious hysteresis on heating and cooling in all these spinel compounds, the peak around 50 K for the composition with $x$ = 4 is related to oxygen contamination [37,38]. However, one feature is that $\chi(T)$, which is primarily consisted of Landau diamagnetism, Larmor diamagnetism, and Pauli paramagnetism, is practically temperature-independent above $T_{MI}$. [17] The magnetic susceptibility at $T_{MI}$ decreases abruptly as temperature decreases as a result of the spin–dimerization transition. Below $T_{MI}$, excepting the low temperatures upturn, the magnetic susceptibility almost maintains constant as shown in **Fig. 4(c)**. $\chi$ values beyond $T_{MI}$ ($\chi_{MI}^{-}$) and below $T_{MI}$ ($\chi_{MI}^{+}$) are given in **Table 3**. Evidently, both $\chi_{MI}^{+}$ and $\chi_{MI}^{-}$ decrease by increasing of the doping amount of S. The magnetic step at $T_{MI}$ is associated to DOS at the Fermi level $N(E_F)$ by the equations:[39] $\Delta\chi = \chi_{MI}^{+} - \chi_{MI}^{-}$, and $\chi_{Pauli} = \frac{3}{2}\Delta\chi = \mu_0\mu_B^2 N(E_F)$, where $\mu_0$ represents the vacuum magneto-conductivity and $\mu_B$ is Bohr magneton, respectively, $\Delta\chi$ is the altitude of magnetic step. The increase of $\Delta\chi$ (see **Table 3**) with increasing $x$ points to the expanding of the bandwidth at the Fermi level caused by the replacement of Te by S. The increase of both $T_{MI}$ and $\Delta\chi$ reveal that the Peierls-like phase transition is enhanced as a result of the lattice reduction. [17] One more obvious feature is that all samples exhibit the magnetization's upturn at low temperatures ascribed to the Curie paramagnetism, [17,39] which may be produced in lattice defects [17,40,41] or paramagnetic impurities.[42,43] Thus, the following formula can be used for fitting the magnetic susceptibility below $T_{MI}$: $\chi = \chi_0 + \frac{C}{T}$, where $\chi_0$ represents the magnetic susceptibility excluding the Curie paramagnetism; $C$ is the Curie parameter. The fitting result on $CuIr_2Te_{0.3}S_{3.7}$ is shown in **Fig. 4(c)** by green solid curves (the fitting data for the other compounds are given in **Fig. S8** in supplemental information). The fitting constants $\chi_0$ and $C$ are given in **Table 3**. Both $\chi_0$ and $C$ decrease by increasing $x$. A previous report shows that the Ag doping in $CuIr_2S_4$ deteriorates the Peierls-like phase transition, [17] therefore weakening the spin–dimerization. Then, for the system $CuIr_2Te_{4-x}S_x$, the number of non-dimerized $Ir^{4+}$ ions will be decreased below $T_{MI}$ reducing the paramagnetism. [17] Hence, as $x$ increases, the paramagnetism becomes weaker resulting in an increase in the Curie constant $C$. The decreased $\chi_0$ may be attributed to the weak remnant ferromagnetism as it has been revealed in different site doped $CuIr_2S_4$. [15,17,18] Correspondingly, the $T_{MI}$ extracted from the magnetic susceptibility (**Fig. 4(b)**, inset) is consistent with the resistivity data. With increasing the S doping concentration $x$ in the range of $3.6 \leq x \leq 4$, the $T_{MI}$ increases gradually. This behavior is similar to that reported in ref. [9] In comparison with the substitution of Se in $CuIr_2S_4$, Te doping is expected to have a robust suppression on the MIT. [9] Nagata *et al.* reported the phase diagram for $CuIr_2(S_{1-x}Se_x)_4$, which displays that the MIT can be kept in a broad substitution range of $0 \leq x \leq 0.7$. [5]

With the aim of calculating the lower critical field ($\mu_0H_{c1}$), the magnetization isotherm $M(H)$ measurements were performed at different fields. **Fig. 5** displays the temperature-dependent $\mu_0H_{c1}$ for the optimal $CuIrTe_{3.85}S_{0.15}$ sample. The bottom inset of **Fig. 5** presents the magnetization $M(H)$ data versus field. The inset at the upper corner of **Fig. 5** displays the full process for estimating $\mu_0H_{c1}$ at different measuring temperatures. To get an accurate $\mu_0H_{c1}$ value, the demagnetization effect should be considered. The demagnetization factor ($N$) values can be estimated according to the equation: $\chi_V = dM/dH$, in which $\chi_V$ denotes the slope of the linearly fitting (see green line in the bottom inset in **Fig. 5**, $N$ value is calculated to be 0.53 ~ 0.63. Then, we can plot the experimental data based on the relationship $M_{Fit} = e + fH$ at low magnetic fields, where $e$ and $f$ represent the intercept and the slope of the linear fitting of the $M(H)$ data, respectively. The relationship of $M\text{-}M_{Fit}$ $(H)$ is plotted in the top inset of **Fig. 5**, which is used to estimate $\mu_0 H_{c1}{}^*$ at the field when $(M\text{-}M_{Fit})$ deviates by ~ 1% below the fitted data ($M$(1%))[44,45]. Subsequently, one can obtain $\mu_0 H_{c1}(T)$ value by using the expression: $\mu_0 H_{c1}(T) = \mu_0 H_{c1}^*(T)/(1-N)$.[46,47] Accordingly, we can fit the $\mu_0 H_{c1}(T)$ values on the basis of the expression: $\mu_0 H_{c1}(T) = \mu_0 H_{c1}(0)[1-(T/T_c)^2]$. Finally, the lower critical field ($\mu_0 H_{c1}(0)$) at zero temperature for the $CuIr_2Te_{3.85}S_{0.15}$ compound is calculated to be 17 mT. Compared to the pristine $CuIr_2Te_4$ compound (28 mT) and Ru doping compound $CuIr_{1.95}Ru_{0.05}Te_4$ (98 mT), [21] the isoelectronic S doping compound has smaller $\mu_0 H_{c1}(0)$s, as summarized in **Table 4**.

Additionally, we analyzed the upper critical field ($\mu_0 H_{c2}$) via the temperature-dependent resistivity measurement under increased applied magnetic fields $\rho(T, H)$ (**Fig. 6**, insets). Here, we calculate the $\mu_0 H_{c2}$ using 50 % criteria of the superconducting transition value from the normalized resistivity ($\rho_N$). It is clear that the $T_c$ decreases upon applying a magnetic field. Consequently, one can calculate the $\mu_0 H_{c2}$ values on the basis of Werthamer-Helfand-Hohenburg (WHH) and Ginzberg-Landau (GL) theories. Then, we can adopt the simplified WHH equation: [48] $\mu_0 H_{c2}(0) = -0.693 T_c (dH_{c2}/dT)|_{T_c}$ to obtain the $\mu_0 H_{c2}(0)$s, [49-53] where $(dH_{c2}/dT_c)$ represents the slope of $\mu_0 H_{c2}(T)$ in the vicinity of $T_c$. Based on the simplified WHH model, we can get the $\mu_0 H_{c2}(0)$ for the $CuIr_2Te_{4-x}S_x$ ($x$ = 0.05, 0.075 and 0.15) compounds, which are 0.140, 0.174 and 0.168 T, correspondingly. These values are all greater than that of the parent $CuIr_2Te_4$. Nonetheless, the highest value of $\mu_0 H_{c2}(0)$ does not correspond to the highest $T_c$. It has been assumed that the obtained $\mu_0 H_{c2}(0)$s for weak-coupling Bardeen-Cooper-Schrieffer (BCS) superconductors are not more than that of the Pauli limiting field ($H^P$ $=1.86{*}T_c$).[54] The values of $H^P$ are calculated to be 4.95, 5.09 and 5.25 T, respectively, which is larger than that of the undoped $CuIr_2Te_4$. Correspondingly, we can further calculate the Ginzburg-Landau coherence length $\xi_{GL}(0)$ from the formula $H_{c2} = \phi_0 / 2\pi\xi_{GL}^2$, using the $H_{c2}(0)$ data from the WHH model, where the flux quantum $\phi_0 = 2.07 \times 10^{-3}$ T $\mu m^2$. The calculated values of $\xi_{GL}(0)$s for $CuIr_2Te_{3.9}S_{0.05}$, $CuIr_2Te_{3.925}S_{0.075}$ and $CuIr_2Te_{3.85}S_{0.15}$ are 47.81, 42.88 and 45.21 nm, respectively. On the other hand, $\mu_0H_{c1}$ is correlated to the coherence length $\xi$ and the magnetic penetration depth $\lambda$ through the relation $\mu_0H_{c1} = (\phi_0/4\pi\lambda^2)\,[\ln(\kappa) + 0.5]$, where $\kappa = \lambda/\xi$ is the *GL* parameter. [55] We can get the magnetic penetration depth $\lambda_{GL}$ = 119 nm using $\mu_0H_{c1}$ = 0.017 T and $\mu_0H_{c2}$ (0) = 0.168 T for the optimal doping level $CuIr_2Te_{3.85}S_{0.15}$, which is slightly higher than that of the undoped $CuIr_2Te_4$ compound (0.12 T), [12] but smaller than that of the optimum Ru-doped $CuIr_2Te_4$ (0.247 T) [13]. We have also calculated the $\mu_0H_{c2}$ values from the *GL* equation:[56] $\mu_0H_{c2}(T) = \mu_0H_{c2}(0)*[1-(T/T_c)^2]/[1+(T/T_c)^2]$, where $T_c$ is taking from the criteria 50 % of $\rho_N$. The $\mu_0H_{c2}(0)$ values from GL model are 0.203, 0.212 and 0.209 T, respectively. As can be seen on the main panels of **Fig. 6**, the $\mu_0H_{c2}$ values from the GL model are higher as compared to the $\mu_0H_{c2}$ values calculated using the WHH model.

The heat capacity measurements deliver more details about the properties of normal and superconducting states. Results of temperature-dependent specific heat capacity at low temperature of the optimal sample $CuIr_2Te_{3.85}S_{0.15}$ are displayed in **Fig. 7**. The data collected under 10 kOe field up to 10 K is well described by $C_p/T$ (T) = $\gamma + \beta T^2$, as presented by the dashed line through the data in **Fig. 7(a)**, where γ and $\beta$ are the constants of electronic specific heat ($C_{el}$) and the lattice coefficient in the phonon contribution ($C_{ph}$), respectively. A sharp anomaly under zero magnetic field is perceived at 2.81 K as shown in **Fig. 7(b).** The fit gives a normal state Sommerfeld coefficient of $\gamma$ = 10.84 mJ mol$^{-1}$ K$^2$ and the lattice coefficients $\beta$ = 3.39 mJ mol$^{-1}$ K$^{-4}$**.** Debye temperature ($\Theta_D$) of about 160 K is calculated from the formula $\Theta_D = (12\pi^4 nR/5\beta)^{1/3}$ where $n$ = 7 is the number of the atoms per formula unit and $R$ is the gas constant. Having $T_c$ and $\Theta_D$, we can gain the electron-phonon coupling coefficient ($\lambda_{ep}$) on the basis of the inverted McMillan expression:[49] $\lambda_{ep} = \frac{1.04 + \mu^* \ln(\Theta_D/1.45T_c)}{((1-0.62\mu^*)\ln(\Theta_D/1.45T_c)-1.04}$, where the repulsive screened coulomb parameter $\mu^*$ is 0.13, which is the commonly used commonly-used with the McMillan equation for metals. [49,57-59] The obtained value of $\lambda_{ep}$ is around 0.65. Having $\gamma$ and $\lambda_{ep}$, the DOS at the Fermi level $N(E_F)$ can be figured out by using the formula $N(E_F) = 3\gamma/(\pi^2 k_B^2(1+\lambda_{ep}))$, where $k_B$ denotes Boltzmann's constant. $N(E_F)$ is found to be 3.17 states/eV per f.u. which is somewhat enhanced compared to that of the host $CuIr_2Te_4$. The increase in $T_c$ and the suppression of CDW in $CuIr_2Te_{3.85}S_{0.15}$ compound can be explained by the enhancement of $N(E_F)$ and it may also be related to the enrichment of the electron–phonon coupling by the S-ion substitution as compared to parent $CuIr_2Te_4$ (see **Table 4**). **Fig. 7(b)** exhibits the electronic contribution to the heat capacity collected at 0 Oe. The obtained $T_c$ for this sample is 2.81 K. The magnitude of the heat capacity jump is estimated to be 1.48, close to the predicted value (1.43) based on the weak coupling BCS theory.

Finally, a rich electronic phase diagram for the $CuIr_2Te_{4-x}S_x$ series is constructed, which features multiple regions separated by the $T_c$, $T_{CDW}$ and $T_{MI}$ versus the doping level $x$, as displayed in **Fig. 8**. Orange regions represent the suppression and re-emergence of the CDW states upon the sulfur substitution. Concretely, the CDW signature in the resistivity disappears with a small S doping content $x$, whereas it reemerges for $0.2 \leq x \leq 0.5$ and is enhanced as $x$ increases in the doping range of 0.2 to 0.5. This phenomenon is similar to the case of single doped $Cu_{1-x}Ag_xIr_{2-y}Zr_yTe_{4-z}(I/Se)_z$, [23, 32, 60, 61] but it differs from the $CuIr_{2-x}(Ru/Ti)_xTe_4$ and $Cu_{0.5-x}Zn_xIrTe_2$ systems without reappearance of CDW transition in the high doping range. [21, 22, 62] Meanwhile, light blue represents the emergence and evolution of SC upon sulfur substitution, in which a small amount S substitution for Te can slightly enhance the $T_c$ and yields the highest $T_c$ of about 2.82 K at $x = 0.15$ for which the improvement of the SC may be due to the enhanced in electron-phonon coupling induced by the S doping, followed by a drop of $T_c$ where the degradation of $T_c$ is may due to the continuous shrinkage of the lattice which is not beneficial for the SC, which was observed in some other reported superconductors.[63, 64] The SC is completely suppressed at $x > 0.4$, leading to a dome-like phase diagram. Light green and purple denote region of the metallic and insulating states, respectively. On the right side of the diagram, the MIT occurs in the whole cubic spinel phase zone ($3.6 \leq x \leq 0.4$) and $T_{MI}$ increases with the S doping content increasing. This suggests that $T_{MI}$ increases with decreasing unit cell volume. With shrinking atomic spacing, one would expect an increase in MIT because the overlap of electron wave functions favors a metallic state. [65]

## Conclusions

In summary, we have synthesized the polycrystalline $CuIr_2Te_{4-x}S_x$ ($0 \leq x \leq 0.4$) solid solutions via a solid state reaction method. A rich electronic phase diagram has been established, which simplifies the rather complicated structural and electrical features in the system $CuIr_2Te_{4-x}S_x$ ($0 \le x \le 0.4$). Altogether, $CuIr_2Te_{4-x}S_x$ ($0 \le x \le 0.4$) stabilize in two types of structures and divided into three zones with a layered trigonal structure for $0 \le x \le 0.5$, cubic spinel structure near the end of their solid solutions ($3.6 \le x \le 4$), and mixed-phase that intermediates between these two regions. Even with the substitution of a small amount of S for Te, CDW can be suppressed. However, the signature of the CDW-like transition can be observed again in the region of $0.2 \le x \le 0.5$ with and increased $T_{CDW}$. On the other hand, S substitution for Te can slightly enhance the $T_c$ and the optimal doping lever is $x = 0.15$ ($CuIr_2Te_{3.85}S_{0.15}$) with the highest $T_c \approx 2.82$ K. The metal-insulator transition exists in the region of $3.6 \le x \le 4$ and $T_{MI}$ is enhanced by S substitution. Based on our results, $CuIr_2Te_{4-x}S_x$ is a potential platform for further study of the interrelationships between different types of electronic orders. Future systematic studies will be important to better understand these interactions and ascertain the physical origin of these electronic instabilities.

**Conflicts of interest**

There are no conflicts to declare.

**Acknowledgements**

The authors acknowledge the financial support by the National Natural Science Foundation of China (11974432), Guangdong Basic and Applied Basic Research Foundation (2019A1515011718), the Fundamental Research Funds for the Central Universities (19lgzd03) and the Pearl River Scholarship Program of Guangdong Province Universities and Colleges (20191001). Meng Wang was supported by the National Nature Science Foundation of China (11904414) and National Key Research and Development Program of China (2019YFA0705702). The work at IOPCAS is supported by the NSFC (12025408, 11921004, 11904391), the National Key R&D Program of China (2018YFA0305702).

**TABLES**

**Table 1** Rietveld refinement structural parameters of $CuIr_2Te_{3.85}S_{0.15}$ with space group *P3-m1* (no. 164) and $CuIr_2Te_{0.1}S_{3.9}$ with space group *Fd*-3*m* (no. 227)

| ***x* = 0.15** | a = b = 3.9378(2) Å and c = 5.3947(3) Å | | | $R_{wp}$ = 6.85% $R_p$ = 3.65%, $R_{exp}$ = 2.11% | |
|---|---|---|---|---|---|
| **Label** | ***x*** | ***y*** | ***z*** | **site** | **Occupancy** |
| Cu | 0 | 0 | 0.5 | 2*b* | 0.5 |
| Ir | 0 | 0 | 0 | 1*a* | 1 |
| Te | 0.3333 | 0.6667 | 0.2529(2) | 2*b* | 0.924(1) |
| S | 0.3333 | 0.6667 | 0.2529(3) | 2*d* | 0.074(1) |
| ***x* = 3.9** | a = b = c = 9.8521(2) Å | | | $R_{wp}$ = 4.2%, $R_p$ = 3.08%, $R_{exp}$ = 2.02%. | |
| **Label** | ***x*** | ***y*** | ***z*** | **site** | **Occupancy** |
| Cu | 0 | 0 | 0 | 8*a* | 1 |
| Ir | 0.625 | 0.6250 | 0.6250 | 16*e* | 1 |
| Te | 0.3879 | 0.3879 | 0.3879(3) | 32*d* | 0.037(2) |
| S | 0.3716 | 0.3716 | 0.3716(1) | 32*d* | 0.963(1) |

**Table 2.** Doping dependent residual resistance ratio (RRR = $R_{300K}/R_{5K}$), superconducting transition temperature ($T_c$), and CDW transition temperature ($T_{CDW}$) for $CuIr_2Te_{4-x}S_x$..

| S amount ($x$) | RRR | $T_c$ (K) | $T_{CDW}$ (K) |
|---|---|---|---|
| 0 | 4.07 | 2.5 | 186 |
| 0.025 | 4.5 | 2.61 | - |
| 0.05 | 4.33 | 2.66 | - |
| 0.075 | 4.33 | 2.74 | - |
| 0.15 | 5.28 | 2.82 | - |
| 0.2 | 2.85 | 2.72 | 110 |
| 0.3 | 3.2 | 2.46 | 138 |
| 0.4 | 2.88 | 2.15 | 166 |
| 0.5 | 2.50 | 1.8 | 217 |

**Table 3.** The magnetic parameters for $CuIr_2Te_{4-x}S_x$.

| S amount ($x$) | $\chi$ $T_{MI}$ ($10^{-4}$ emu / mol) | | $\Delta\chi$ | $\chi = \chi_0 + C/T$ | |
|---|---|---|---|---|---|
| | $X_{TMI+}$ | $X_{TMI-}$ | | $\chi_0$ ($10^{-4}$ emu / mol) | $C$ (emu K / mol) |
| 3.6 | 1.29 | 0.04 | 1.25 | -0.312 | 0.0130 |
| 3.7 | 0.97 | -0.37 | 1.33 | -0.435 | 0.0023 |
| 3.8 | 0.90 | -0.48 | 1.38 | -0.588 | 0.0031 |
| 3.9 | 0.85 | -0.57 | 1.42 | -0.620 | 0.0016 |
| 4 | 0.69 | -0.80 | 1.49 | -0.770 | 0.0048 |

**Table 4.** Superconducting parameters of different telluride chalcogenides compounds.

| Parameter \ Material | $CuIr_2Te_{3.95}S_{0.05}$ | $CuIr_2Te_{3.925}S_{0.075}$ | $CuIr_2Te_{3.85}S_{0.15}$ | $CuIr_2Te_4$ [20] | $CuIr_{1.95}Ru_{0.05}Te_4$ [21] | $CuIr_2Te_{3.9}I_{0.1}$ [23] | $Cu_{0.25}$ $Zn_{0.25}IrTe_2$ [22] | $CuIr_{1.95}Ti_{0.05}Te_4$ [59] | $CuIr_2Te_{3.9}Se_{0.1}$[32] | $Cu_{0.88}$ $Ag_{0.12}Ir_2Te_4$[57] |
|---|---|---|---|---|---|---|---|---|---|---|
| $T_c$ (K) | 2.66 | 2.74 | 2.82 | 2.5 | 2.79 | 2.95 | 2.82 | 2.84 | 2.83 | 2.93 |
| $\gamma$ (mJ mol$^{-1}$ K$^{-2}$) | | | 10.83 | 12.05 | 12.26 | 12.97 | 13.37 | 14.13 | 10.84 | 13.9 |
| $ß$(mJ mol-1 K$^{-4}$) | | | 3.39 | 1.97 | 1.87 | 3.03 | 1.96 | 2.72 | 3.51 | 2.12 |
| $\Theta_D$ (K) | | | 158 | 190 | 193 | 165 | 190.6 | 170.9 | 157 | 186 |
| $\Delta C/\gamma T_c$ | | | 1.48 | 1.5 | 1.51 | 1.46 | 1.45 | 1.34 | 1.51 | 1.44 |
| $\lambda_{ep}$ | | | 0.68 | 0.63 | 0.65 | 0.70 | 0.66 | 0.64 | 0.65 | 0.64 |
| $N(E_F)$ (states/eV f.u) | | | 3.17 | 3.1 | 3.15 | 3.24 | 3.41 | 3.67 | 3.11 | 3.61 |
| $\mu H_{c1}(0)$ (mT) | | | 17 | 28 | 98 | 24 | 62 | 95 | 66 | 13.5 |
| $\mu H_{c2}(0)$ (T) (G-L) | 0.203 | 0.212 | 0.209 | 0.145 | | 0.232 | 0.198 | | 0.148 | |
| $\mu H_{c2}(0)$(T)(WHH) | 0.140 | 0.174 | 0.168 | 0.12 | 0.247 | 0.188 | | | 0.144 | 0.21 |
| $-dHc_2/dT_c$ (T/K) | 0.076 | 0.092 | 0.086 | 0.066 | 0.125 | | | | 0.073 | |
| $\mu_0^{Hp}$ (T) | 4.95 | 5.09 | 5.25 | 4.65 | 5.24 | 5.49 | 5.26 | 5.28 | 5.26 | 5.2 |
| $\xi_{GL}$(nm) | 47.81 | 42.88 | 45.21 | 52.8 | 36.3 | 41.9 | 40.7 | | 47.18 | 40 |

## FIGURE LEGENDS

**Fig. 1.** (Color online) Rietveld refinements for the representative samples (a) $CuIr_2Te_{3.85}S_{0.15}$ and (b) $CuIr_2Te_{0.1}S_{3.9}$. The insets show the crystallographic structures of $CuIr_2Te_{4-x}S_x$ compounds.

**Fig. 2.** (Color online) **(a-b)** Room temperature PXRD patterns for $CuIr_2Te_{4-x}S_x$ ($0 \le x \le 4$). (c) The variation of unit-cell constants *a* and *c* with *x* content. Blue hollow circle stands for *c*, and red hollow circle notes for *a*.

**Fig. 3.** (Color online) Transport and magnetization characterizations for $CuIr_2Te_{4-x}S_x$. **(a)** The resistivity measurements as a function of temperature for polycrystalline $CuIr_2Te_{4-x}S_x$ series ($0 \le x \le 0.5$). **(b)** The resistivity ratio ($\rho/\rho_{300\text{ K}}$) as a function of temperature for the polycrystalline $CuIr_2Te_{4-x}S_x$ series at low temperatures ($0 \le x \le 0.5$), showing the superconducting transition temperatures. **(c)** Magnetization curves for $CuIr_2Te_{4-x}S_x$ ($0 \le x \le 0.5$) at low temperatures under 30 Oe applied fields, marking the onset of the superconducting transition temperatures. **(d)** Magnetization curves under applied field H 10 kOe for polycrystalline $CuIr_2Te_{4-x}S_x$ ($0.2 \le x \le 0.5$).

**Fig. 4** (Color online) (a) Temperature-dependent resistivity for the cubic spinel samples of $CuIr_2Te_{4-x}S_x$ ($3.6 \le x \le 4$). The close circles are for cooling and the open ones are for warming **(b)** Magnetization measurements as a function of temperature for the spinel samples of $CuIr_2Te_{4-x}S_x$ ($3.9 \le x \le 4$) measured under applied field H = 10 kOe. The insets of **Fig. 4 (a, b)** show the amplified plots near the MIT). (c) The temperature dependence of magnetic susceptibility for $CuIr_2Te_{0.3}S_{3.7}$. The magnetic susceptibility between 4 K and $T_{MI}$ are fitted by $T_{MI}$: $\chi = \chi_0 + \frac{C}{T}$ (green solid line).

**Fig. 5** (Color online) The lower critical fields for $CuIr_2Te_{3.85}S_{0.15}$, with the fitting lines using the equation $\mu_0 H_{c1}(T) = \mu_0 H_{c1}(0)[1-(T/T_c)^2]$. The top insets display the magnetic susceptibilities M(H) curves at different temperatures. The bottom insets display $M$-$M_{\text{Fit}}$ ($H$) vs. temperature ($T$).

**Fig. 6.** (Color online) (**a-c**) The temperature dependence of upper critical fields curves for $CuIr_2Te_{3.95}S_{0.05}$, $CuIr_2Te_{3.925}S_{0.075}$ and $CuIr_2Te_{3.85}S_{0.15}$ respectively. The data are fitted using WHH (the color solid lines) and GL (the color dashed lines) models. the insets depict the corresponding resistivity measurements as a function of temperature under different magnetic field $\rho(T,H)$

**Fig. 7.** (Color online) **(a)** Low temperature specific heat capacity ($C_p/T(T)$) at 0 Oe (green solid circles) and 10 kOe field (red circles) as a function of $T^2$, the inset shows heat capacity jump. **(b)** The electronic portion of the heat capacity ($C_{el}/T$) vs $T$.

**Fig. 8** (Color online) The electronic phase diagram for $CuIr_2Te_{4-x}S_x$ ($0 \leq x \leq 0.4$). $T_{CDW}$ and $T_{MI}$ are identified from the cooling mode of $\rho(T)$ *and* $\chi(T)$.

**Fig. 1**

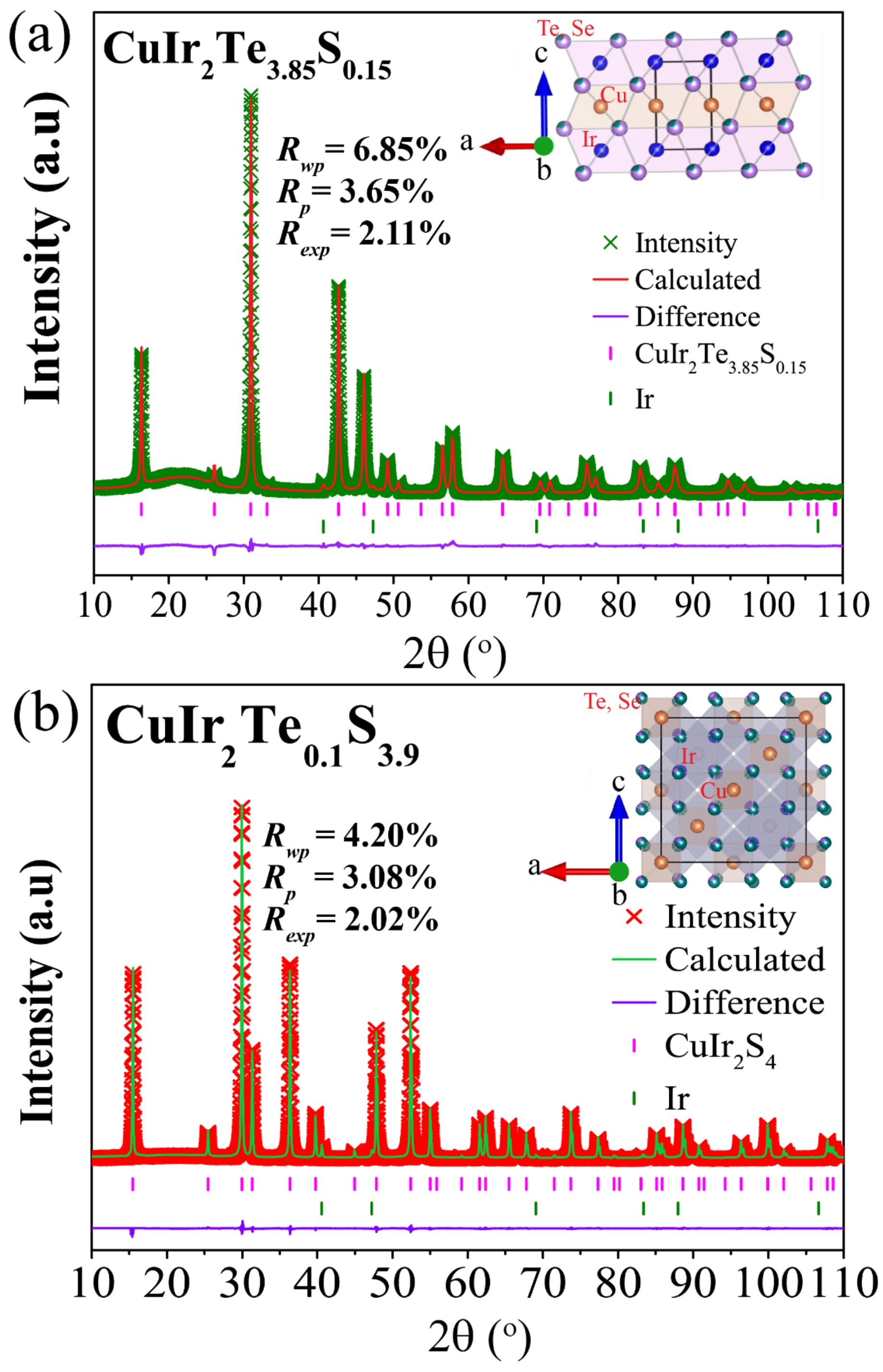

(a)
CuIr2Te3.85S0.15
Rwp = 6.85%
Rp = 3.65%
Rexp = 2.11%
Te, Se
Cu
Ir
c
a
b
Intensity
Calculated
Difference
CuIr2Te3.85S0.15
Ir
Intensity (a.u)
10 20 30 40 50 60 70 80 90 100 110
2θ (°)
(b)
CuIr2Te0.1S3.9
Rwp = 4.20%
Rp = 3.08%
Rexp = 2.02%
Te, Se
Ir
Cu
c
a
b
Intensity
Calculated
Difference
CuIr2S4
Ir
Intensity (a.u)
10 20 30 40 50 60 70 80 90 100 110
2θ (°)

**Fig. 2**

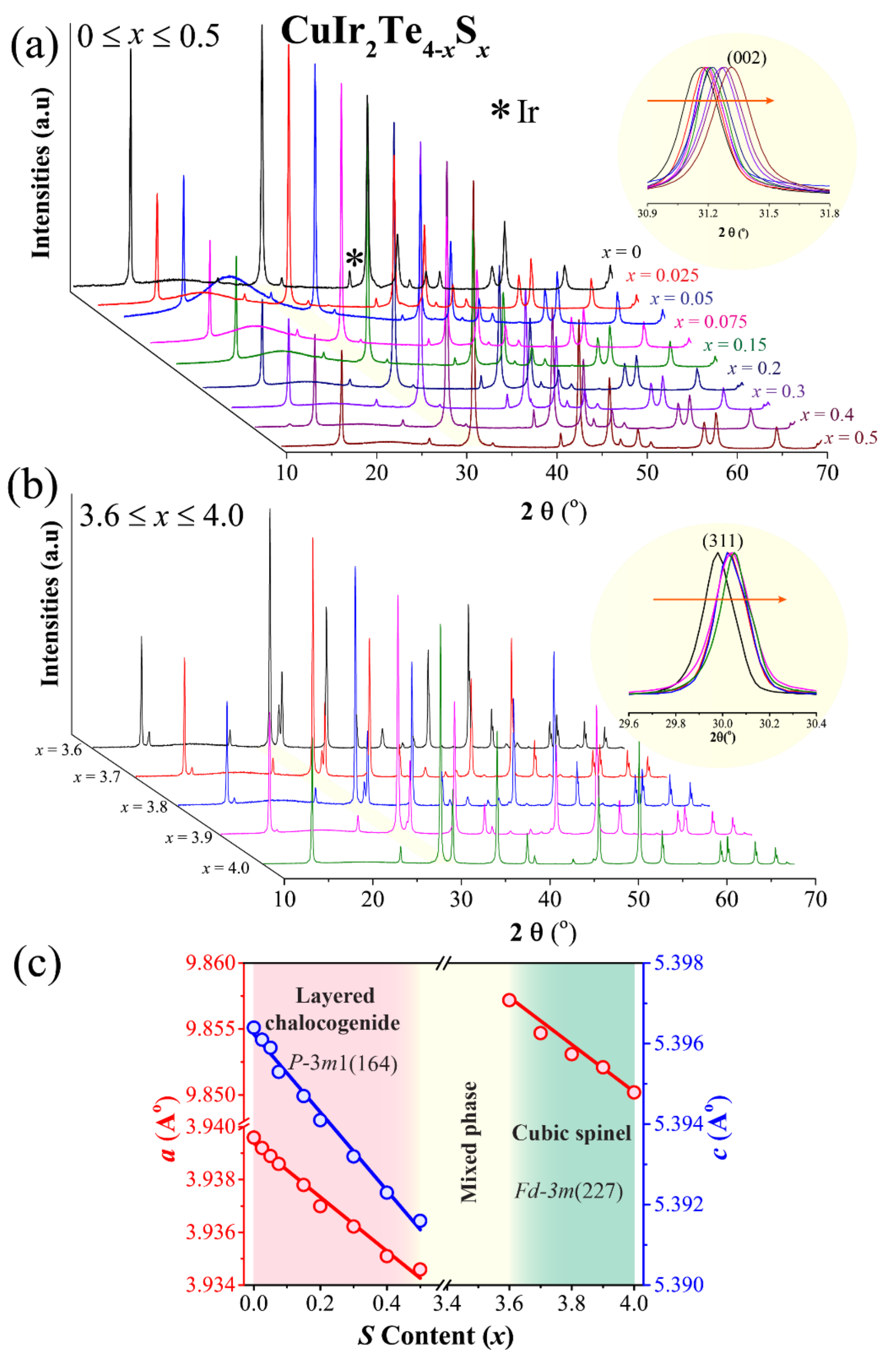

**Fig. 3**

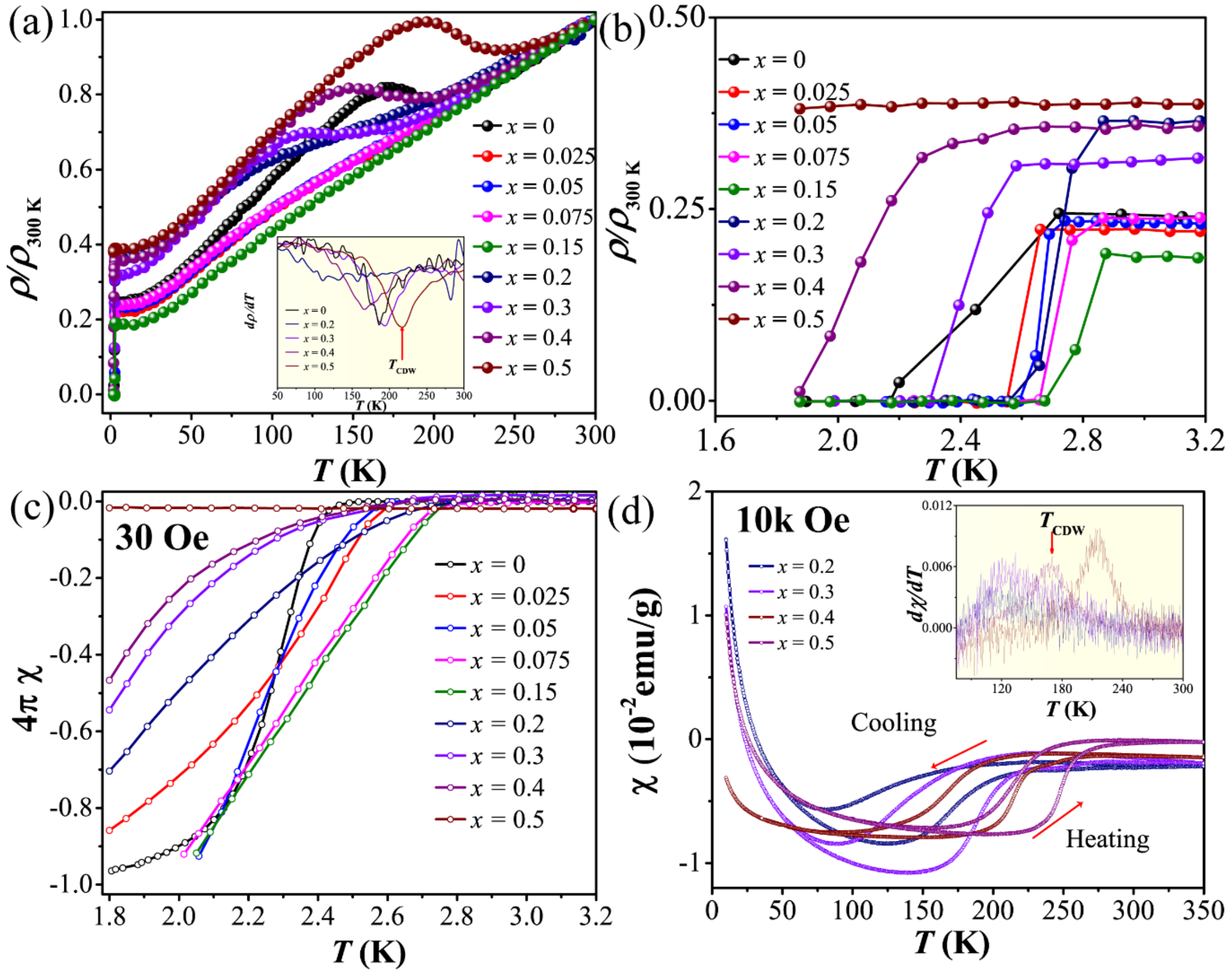

**Fig. 4**

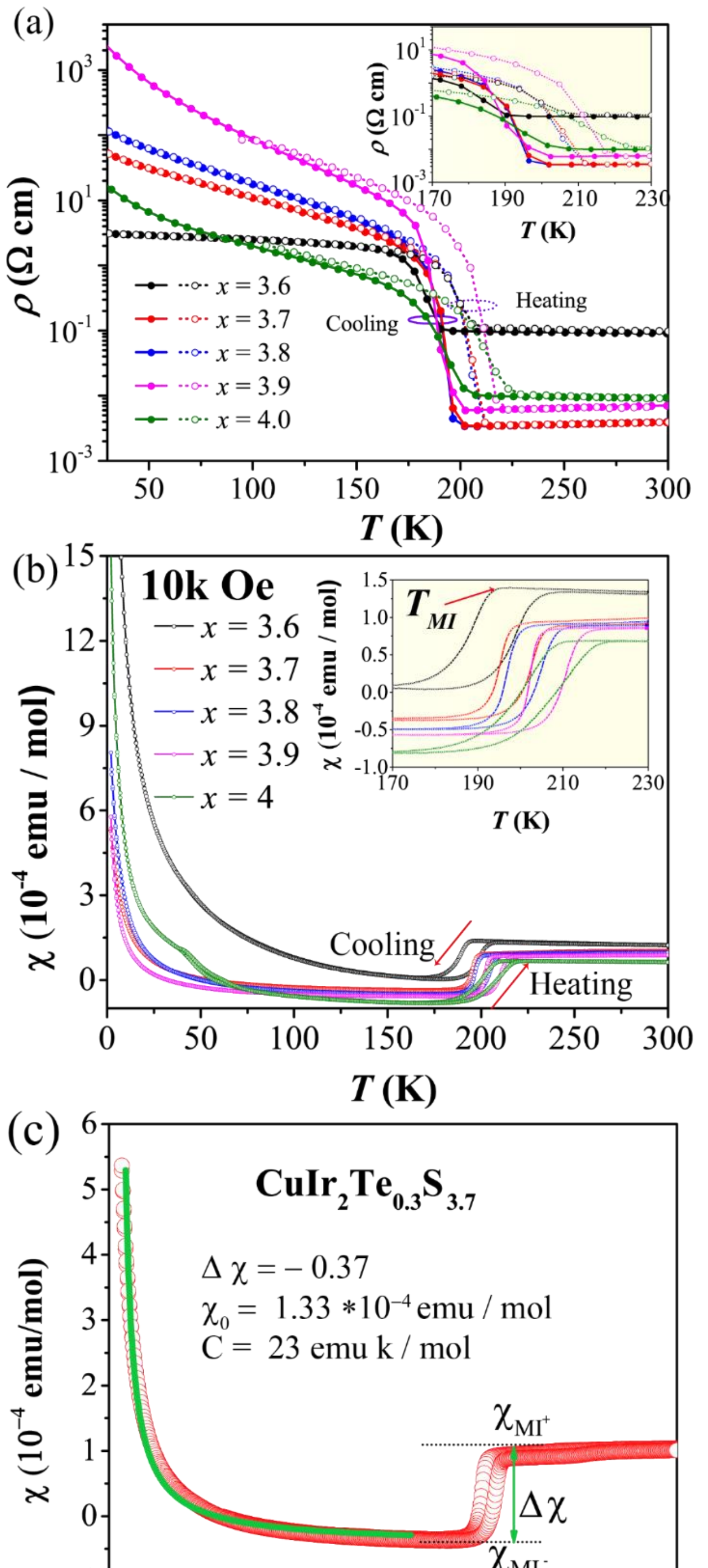

Fig. 5

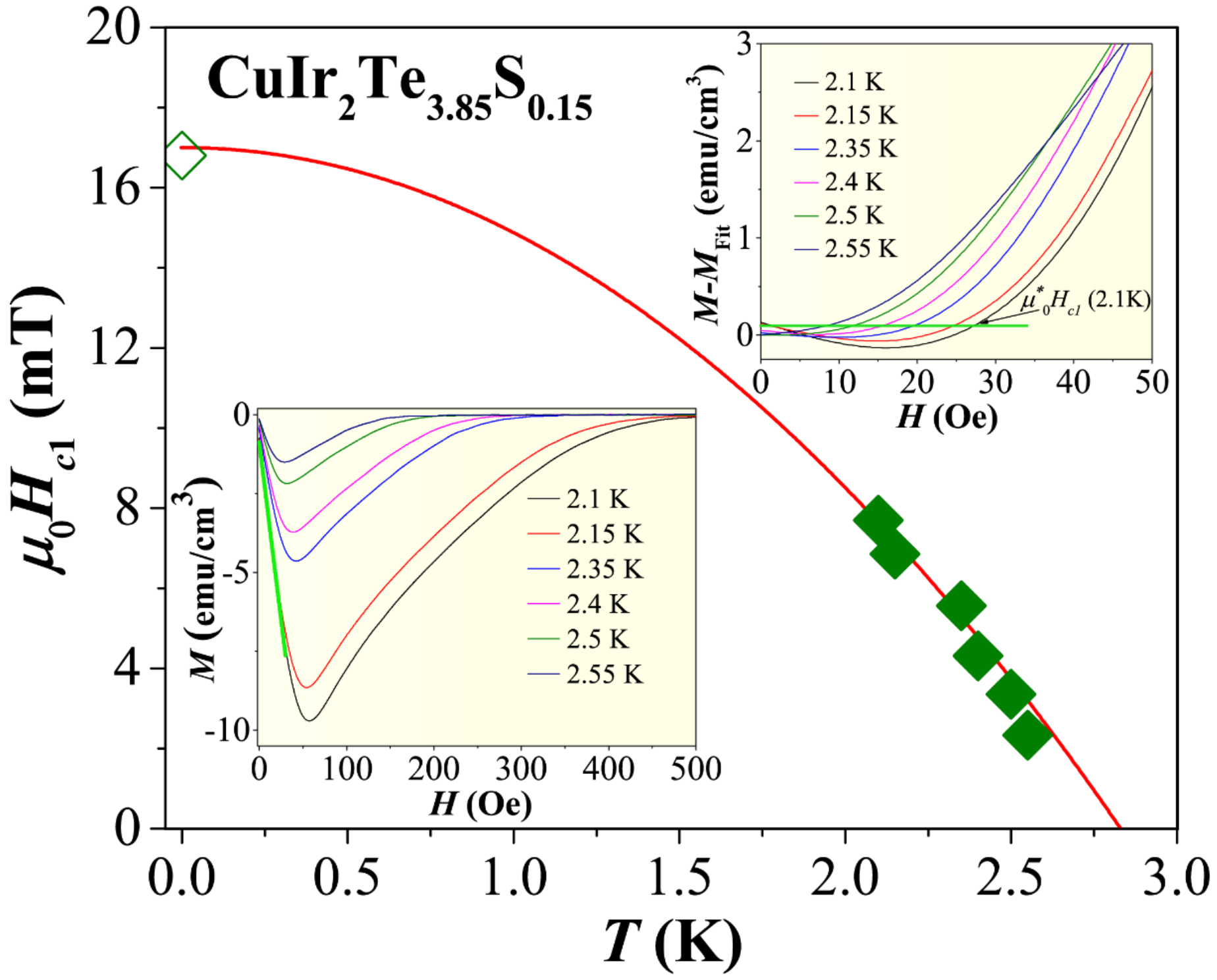

**Fig. 6**

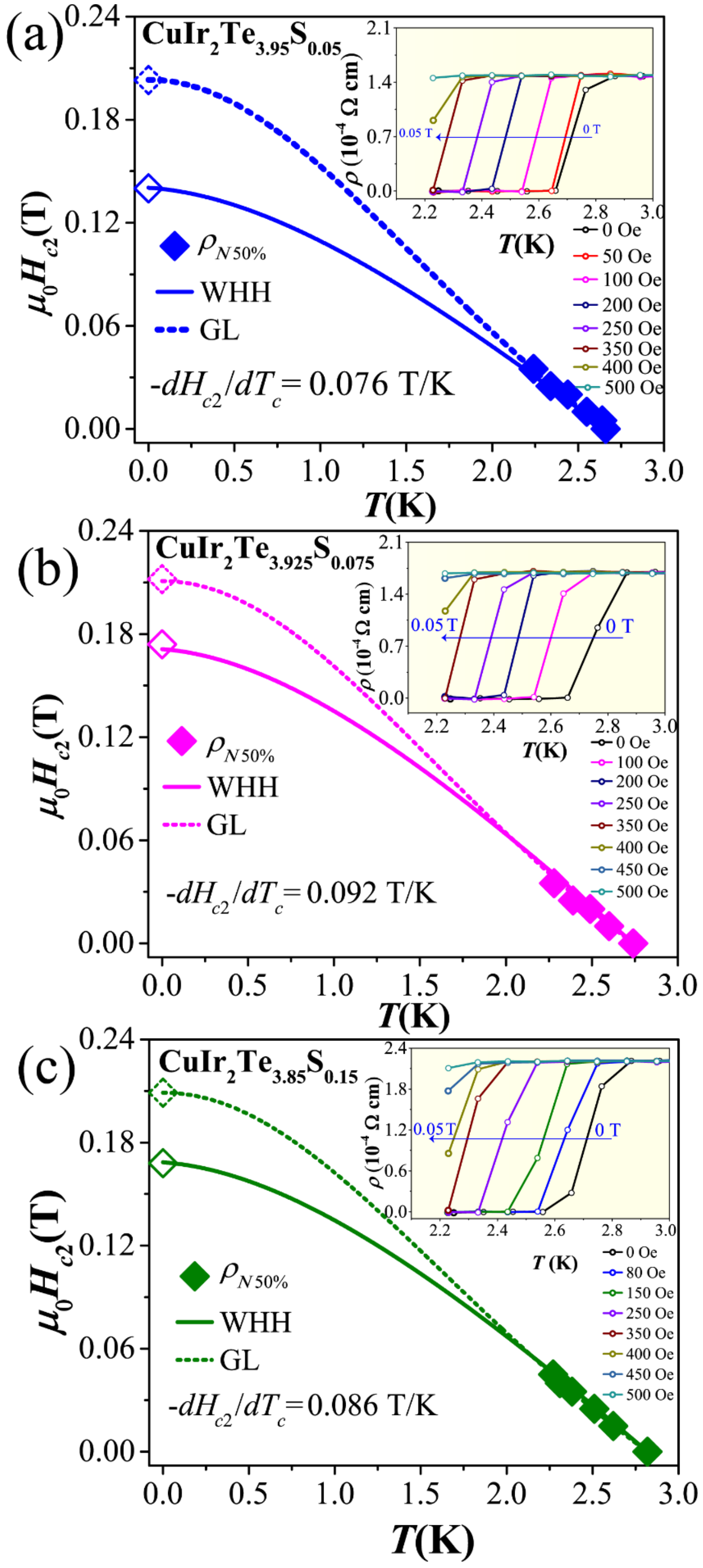

**Fig. 7**

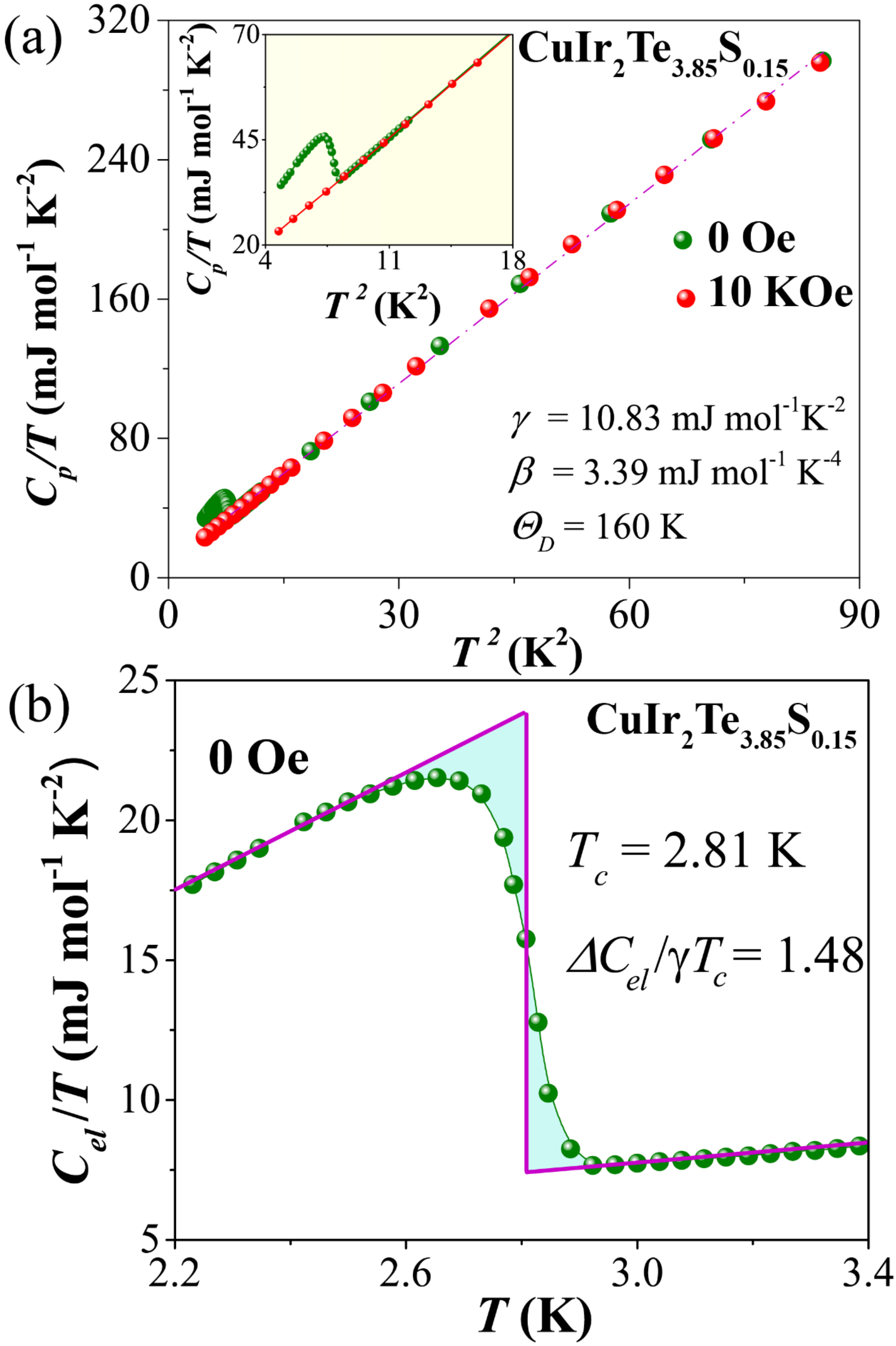

**Fig. 8**

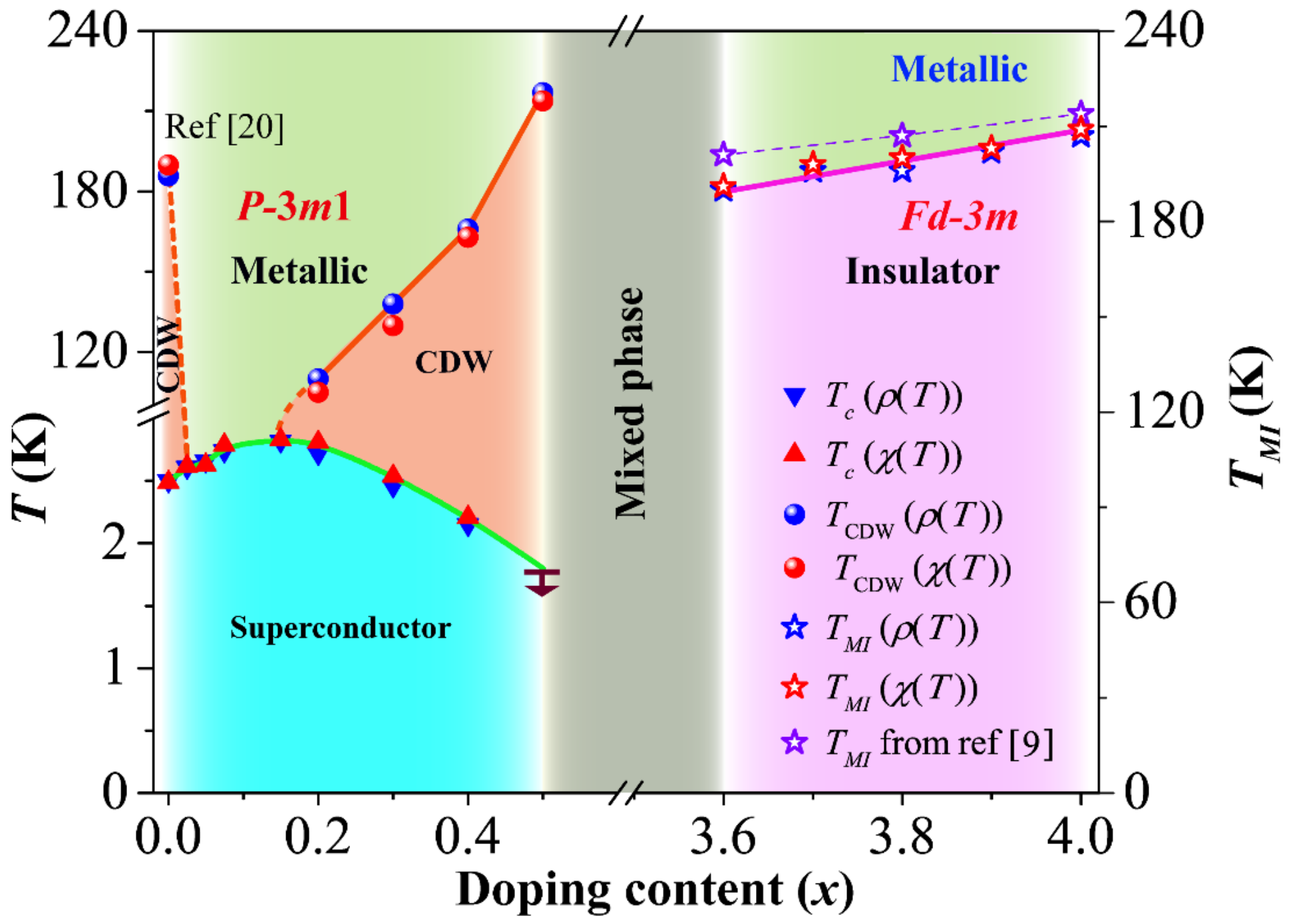

# Supplemental Information

## Superconducting dome associated with the suppression and reemergence of charge-density-wave states upon sulfur substitution in the $CuIr_2Te_{4-x}S_x$ chalcogenides

*Mebrouka Boubeche[a], Ningning Wang[b], Jianping Sun[b], Xiping Zhu[b], Lingyong Zeng[a], Shaojuan Luo[c], Yiyi He[a], Jia Yu[d], Meng Wang[d], Jinguang Cheng[b], Huixia Luo[a*]*

[a] *School of Materials Science and Engineering, State Key Laboratory of Optoelectronic Materials and Technologies, Key Lab of Polymer Composite & Functional Materials, Guangzhou Key Laboratory of Flexible Electronic Materials and Wearable Devices, Sun Yat-Sen University, No. 135, Xingang Xi Road, Guangzhou, 510275, P. R. China*

[b]*Beijing National Laboratory for Condensed Matter Physics and Institute of Physics, Chinese Academy of Sciences and School of Physical Sciences, University of Chinese Academy of Sciences, Beijing 100190, China*

[c]*School of Chemical Engineering and Light Industry, Guangdong University of Technology, Guangzhou,* 510006, *P. R. China*

[d] *Center for Neutron Science and Technology, School of Physics, Sun Yat-Sen University, Guangzhou, 510275, China*

*E-mail address: luohx7@mail.sysu.edu.cn;*

[*]*Corresponding author/authors complete details (Telephone; E-mail:) (+86)-2039386124*

**Table S1**. Rietveld refinement structural parameters of $CuIr_2Te_{4-x}S_x$ compounds series.

| ***x* = 0.025** | *a* = *b* = 3.9392(3) Å and *c* = 5.3964(3) Å | | | $R_{wp}$ = 5.82% $R_p$ = 3.24%, $R_{exp}$ = 2.09% | |
|---|---|---|---|---|---|
| **Label** | ***x*** | ***y*** | ***Z*** | **site** | **Occupancy** |
| Cu | 0 | 0 | 0.5 | 2*b* | 1 |
| Ir | 0 | 0 | 0 | 1*a* | 1 |
| Te | 0.3333 | 0.6667 | 0.2532(1) | 2*b* | 0.985(1) |
| S | 0.3333 | 0.6667 | 0.2532(3) | 2*d* | 0.012(1) |
| ***x* = 0.05** | *a* = *b* = 3.9389 (1) Å and *c* = 5.3961(2) Å | | | $R_{wp}$ = 5.31%, $R_p$ = 3.50%, $R_{exp}$ = 2.10%. | |
| **Label** | ***x*** | ***y*** | ***z*** | **site** | **Occupancy** |
| Cu | 0 | 0 | 0.5 | 2*b* | 1 |
| Ir | 0 | 0 | 0 | 1*a* | 1 |
| Te | 0.3333 | 0.6667 | 0.2529(2) | 2*b* | 0.971(3) |
| S | 0.3333 | 0.6667 | 0.2529(2) | 2*d* | 0.024(2) |
| ***x* = 0.075** | *a* = *b* = 3.9383(4) Å and *c* = 5.3953(2) Å | | | $R_{wp}$ =5.78%, $R_p$ = 3.08%, $R_{exp}$ = 2.04%. | |
| **Label** | ***x*** | ***y*** | ***z*** | **Site** | **Occupancy** |
| Cu | 0 | 0 | 0.5 | 2*b* | 1 |
| Ir | 0 | 0 | 0 | 1*a* | 1 |
| Te | 0.3333 | 0.6667 | 0.2525(4) | 2*b* | 0.967(3) |
| S | 0.3333 | 0.6667 | 0.2525(4) | 2*d* | 0.032(3) |
| ***x* = 0.2** | *a* = *b* = 3.937 (3) Å and *c* = 5.3941(7) Å | | | $R_{wp}$ = 5.91%, $R_p$ = 3.2%, $R_{exp}$ = 2.13%. | |
| **Label** | ***x*** | ***y*** | ***z*** | **Site** | **Occupancy** |
| Cu | 0 | 0 | 0.5 | 2*b* | 1 |
| Ir | 0 | 0 | 0 | 1*a* | 1 |
| Te | 0.3333 | 0.6667 | 0.2522(3) | 2*b* | 0.900(1) |
| S | 0.3333 | 0.6667 | 0.2522(3) | 2*d* | 0.098(2) |
| ***x* = 0.3** | *a* = *b* = 3.9362 (4) Å and *c* = 5.3932(5) Å | | | $R_{wp}$ = 6.02%, $R_p$ = 3.20%, $R_{exp}$ = 2.12%. | |
| **Label** | ***x*** | ***y*** | ***z*** | **site** | **Occupancy** |
| Cu | 0 | 0 | 0.5 | 2*b* | 1 |
| Ir | 0 | 0 | 0 | 1*a* | 1 |
| Te | 0.3333 | 0.6667 | 0.2520(3) | 2*b* | 0.850(4) |
| S | 0.3333 | 0.6667 | 0.2520(3) | 2*d* | 0.024(3) |
| ***x* = 0.4** | *a* = *b* = 3.9351 (5) Å and *c* = 5.3922(7) Å | | | $R_{wp}$ = 6.1%, $R_p$ = 3.17%, $R_{exp}$ = 2.15%. | |
| Cu | 0 | 0 | 0.5 | 2*b* | 1 |
| Ir | 0 | 0 | 0 | 1*a* | 1 |
| Te | 0.3333 | 0.6667 | 0.2518(6) | 2*b* | 0.80(2) |
| S | 0.3333 | 0.6667 | 0.2518(6) | 2*d* | 0.20(3) |
| ***x* = 0.5** | *a* = *b* = 3.9351 (5) Å and *c* = 5.3922(7) Å | | | $R_{wp}$ = 4.2%, $R_p$ = 3.08%, $R_{exp}$ = 2.22% | |
| **Label** | ***x*** | ***y*** | ***Z*** | **site** | **Occupancy** |
| Cu | 0 | 0 | 0.5 | 2*b* | 1 |
| Ir | 0 | 0 | 0 | 1*a* | 1 |
| Te | 0.3333 | 0.6667 | 0.2515(4) | 2*b* | 0.75(6) |
| S | 0.3333 | 0.6667 | 0.2515(4) | 2*d* | 0.25(4) |
| ***x* = 3.6** | *a* = *b* = *c* = 9.8572(7) Å | | | $R_{wp}$ = 6.2%, $R_p$ = 4.02%, $R_{exp}$ = 2.34% | |
| **Label** | ***x*** | ***y*** | ***Z*** | **site** | **Occupancy** |
| Cu | 0 | 0 | 0 | 8*a* | 1 |
| Ir | 0.625 | 0.6250 | 0.6250 | 16*e* | 1 |
| S | 0.3716 | 0.3716 | 0.3712(3) | 32*d* | 0.852(5) |
| Te | 0.3879 | 0.3879 | 0.3872(4) | 32*d* | 0.148(4) |
| ***x* = 3.7** | *a* = *b* = *c* = 9.8547 (5) Å | | | $R_{wp}$ = 5.89%, $R_p$ = 3.75%, $R_{exp}$ = 2.25% | |
| **Label** | ***x*** | ***y*** | ***z*** | **Site** | **Occupancy** |
| Cu | 0 | 0 | 0 | 8*a* | 1 |
| Ir | 0.625 | 0.6250 | 0.6250 | 16*e* | 1 |
| S | 0.3716 | 0.3716 | 0.3704(5) | 32*d* | 0.890(4) |
| Te | 0.3879 | 0.3879 | 0.3861(5) | 32*d* | 0.110(2) |
| ***x* = 3.8** | *a* = *b* = *c* = 9.8531(2) Å | | | $R_{wp}$ = 4.15%, $R_p$ = 3.15%, $R_{exp}$ = 2.1% | |
| **Label** | ***x*** | ***y*** | ***Z*** | **site** | **Occupancy** |
| Cu | 0 | 0 | 0 | 8*a* | 1 |
| Ir | 0.625 | 0.6250 | 0.6250 | 16*e* | 1 |
| S | 0.3716 | 0.3716 | 0.3608(6) | 32*d* | 0.926(5) |
| Te | 0.3879 | 0.3879 | 0.3856(2) | 32*d* | 0.074(6) |
| ***x* = 4** | $a = b = c = 9.8502(2)$ Å | | | $R_{wp} = 3.8\%$, $R_p = 2.92\%$, $R_{exp} = 2.02\%$ | |
| **Label** | ***x*** | ***y*** | ***Z*** | **site** | **Occupancy** |
| Cu | 0 | 0 | 0 | 8*a* | 1 |
| Ir | 0.625 | 0.6250 | 0.6250 | 16*e* | 1 |
| S | 0.3879 | 0.3879 | 0.3839(7) | 32*d* | 1 |

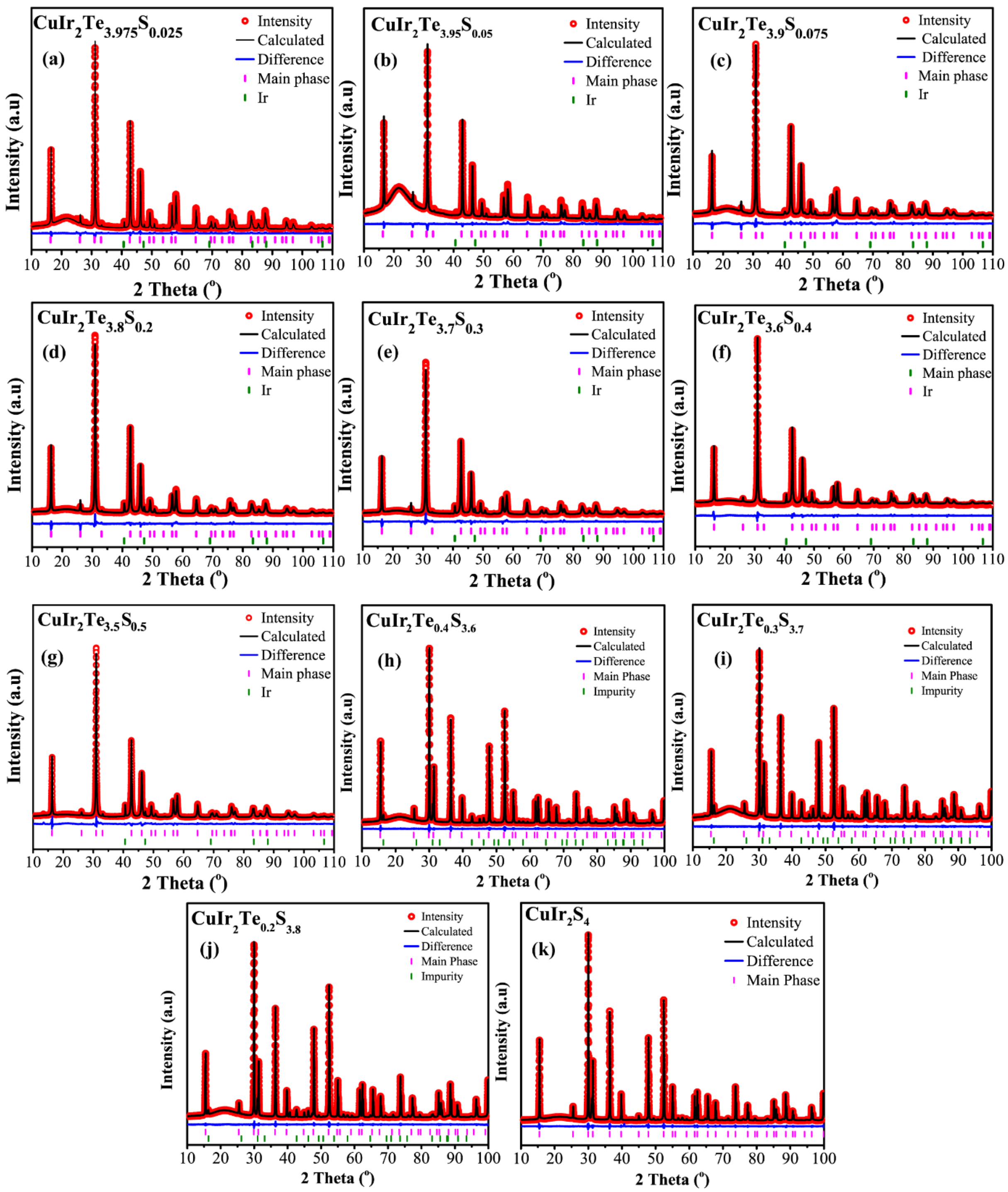

**Fig. S1. (a-k)** Rietveld refinement graphs of $CuIr_2Te_{4-x}S_x$ polycrystalline specimens.

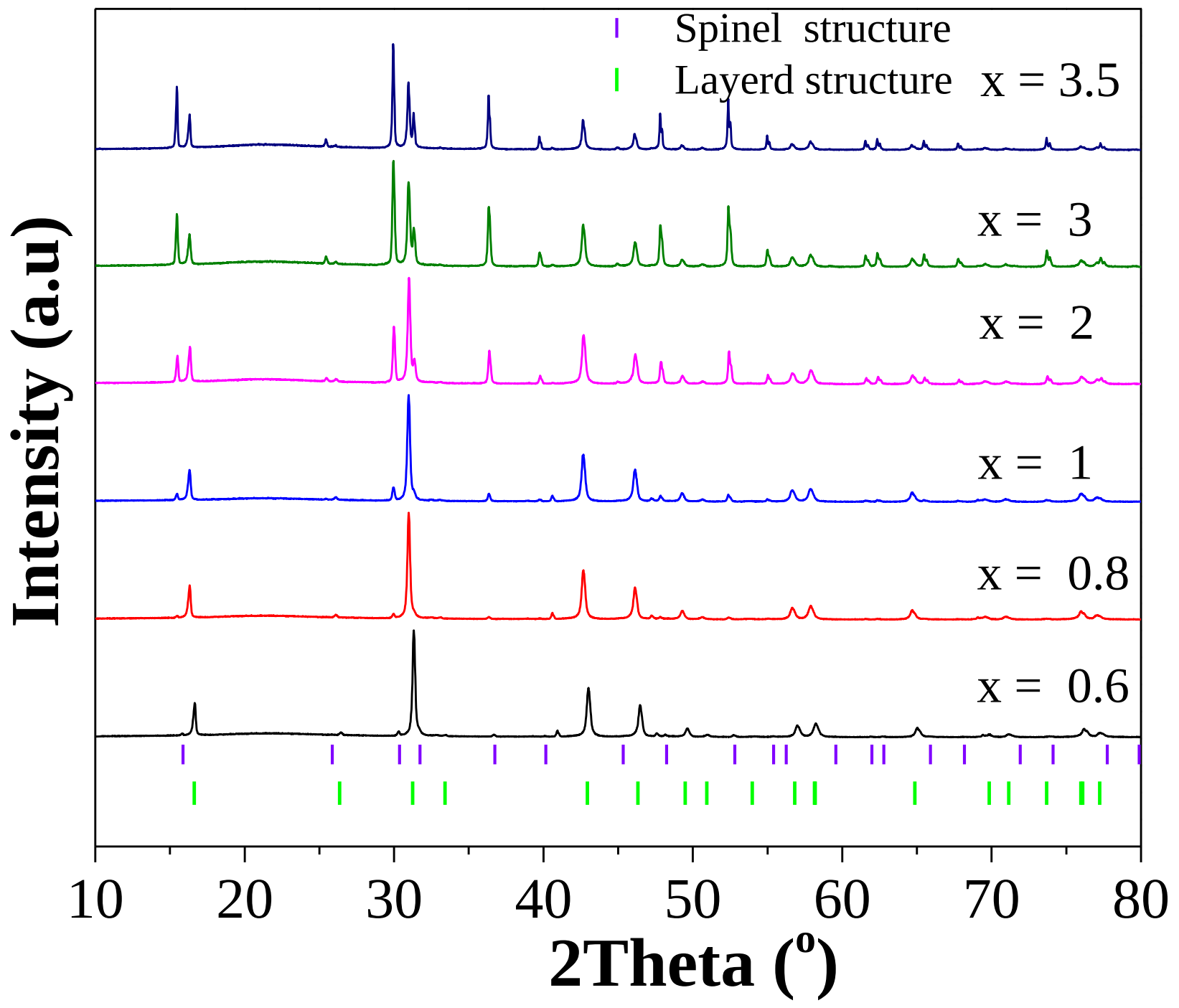

**Fig. S2.** XRD patterns of $CuIr_2Te_{4-x}S_x$ ($0.5 < x < 3.6$) polycrystalline specimens.

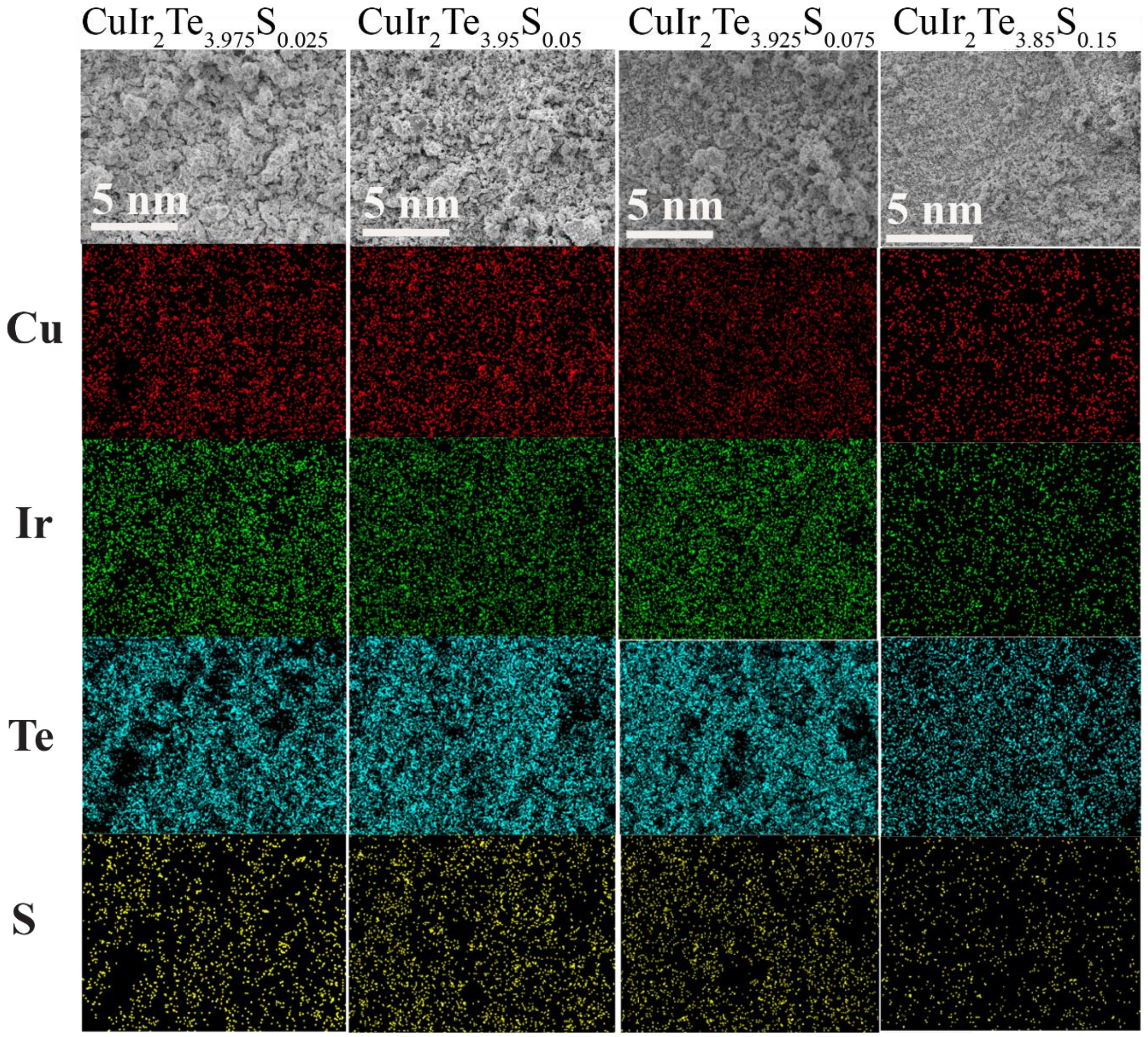

**Fig. S3.** SEM pictures and EDXS mappings for the elements in the light doping $CuIr_2Te_{4-x}S_x$ ($x$ = 0.025, 0.05, 0.075, 0.15) powders.

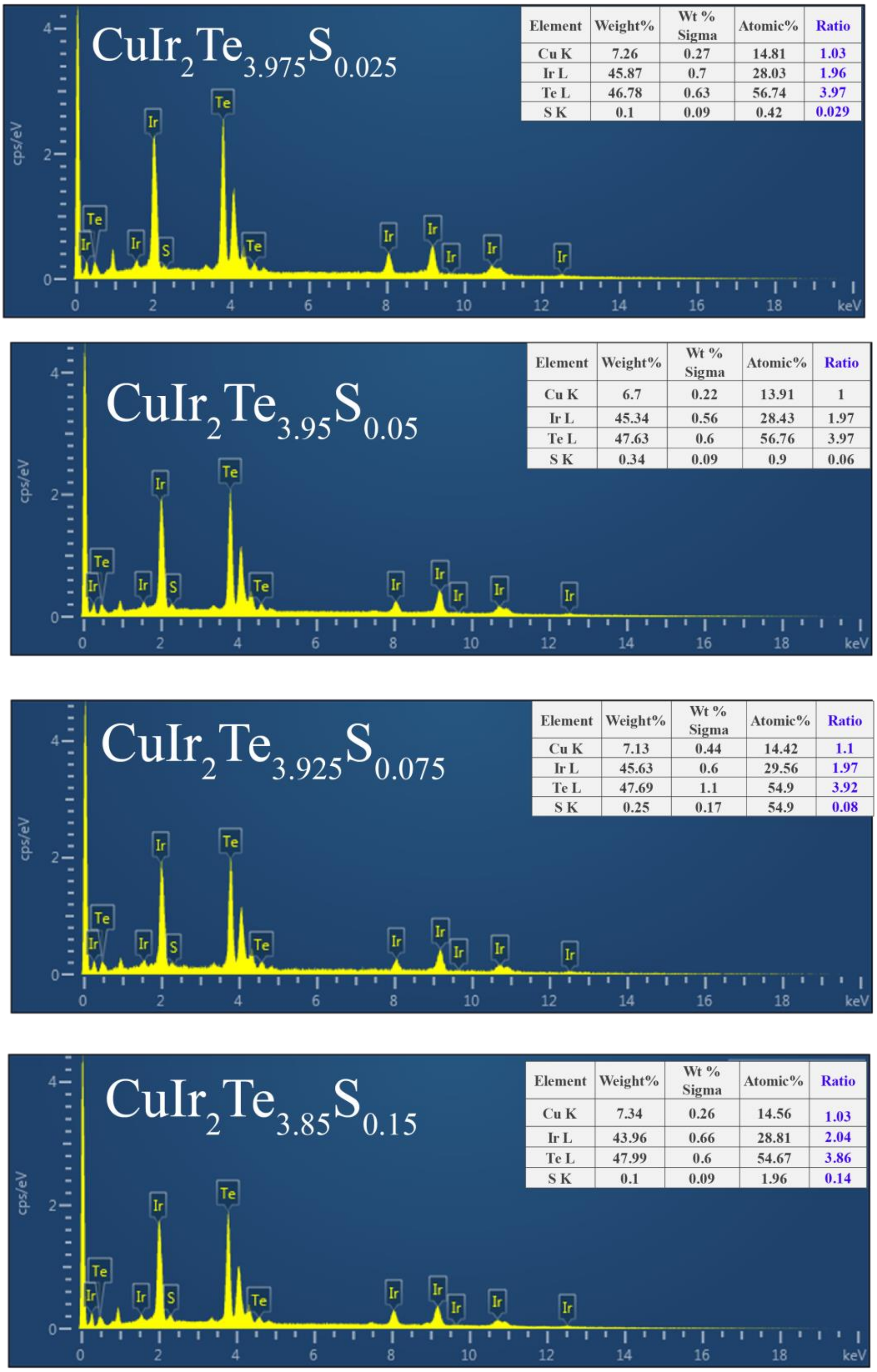

| Element | Weight% | Wt % Sigma | Atomic% | Ratio |
|---|---|---|---|---|
| Cu K | 7.26 | 0.27 | 14.81 | 1.03 |
| Ir L | 45.87 | 0.7 | 28.03 | 1.96 |
| Te L | 46.78 | 0.63 | 56.74 | 3.97 |
| S K | 0.1 | 0.09 | 0.42 | 0.029 |

| Element | Weight% | Wt % Sigma | Atomic% | Ratio |
|---|---|---|---|---|
| Cu K | 6.7 | 0.22 | 13.91 | 1 |
| Ir L | 45.34 | 0.56 | 28.43 | 1.97 |
| Te L | 47.63 | 0.6 | 56.76 | 3.97 |
| S K | 0.34 | 0.09 | 0.9 | 0.06 |

| Element | Weight% | Wt % Sigma | Atomic% | Ratio |
|---|---|---|---|---|
| Cu K | 7.13 | 0.44 | 14.42 | 1.1 |
| Ir L | 45.63 | 0.6 | 29.56 | 1.97 |
| Te L | 47.69 | 1.1 | 54.9 | 3.92 |
| S K | 0.25 | 0.17 | 54.9 | 0.08 |

| Element | Weight% | Wt % Sigma | Atomic% | Ratio |
|---|---|---|---|---|
| Cu K | 7.34 | 0.26 | 14.56 | 1.03 |
| Ir L | 43.96 | 0.66 | 28.81 | 2.04 |
| Te L | 47.99 | 0.6 | 54.67 | 3.86 |
| S K | 0.1 | 0.09 | 1.96 | 0.14 |

**Fig. S4.** EDXS spectra and element ratios for the light doping $CuIr_2Te_{4-x}S_x$ ($x$ = 0.025, 0.05, 0.075, 0.15) powders.

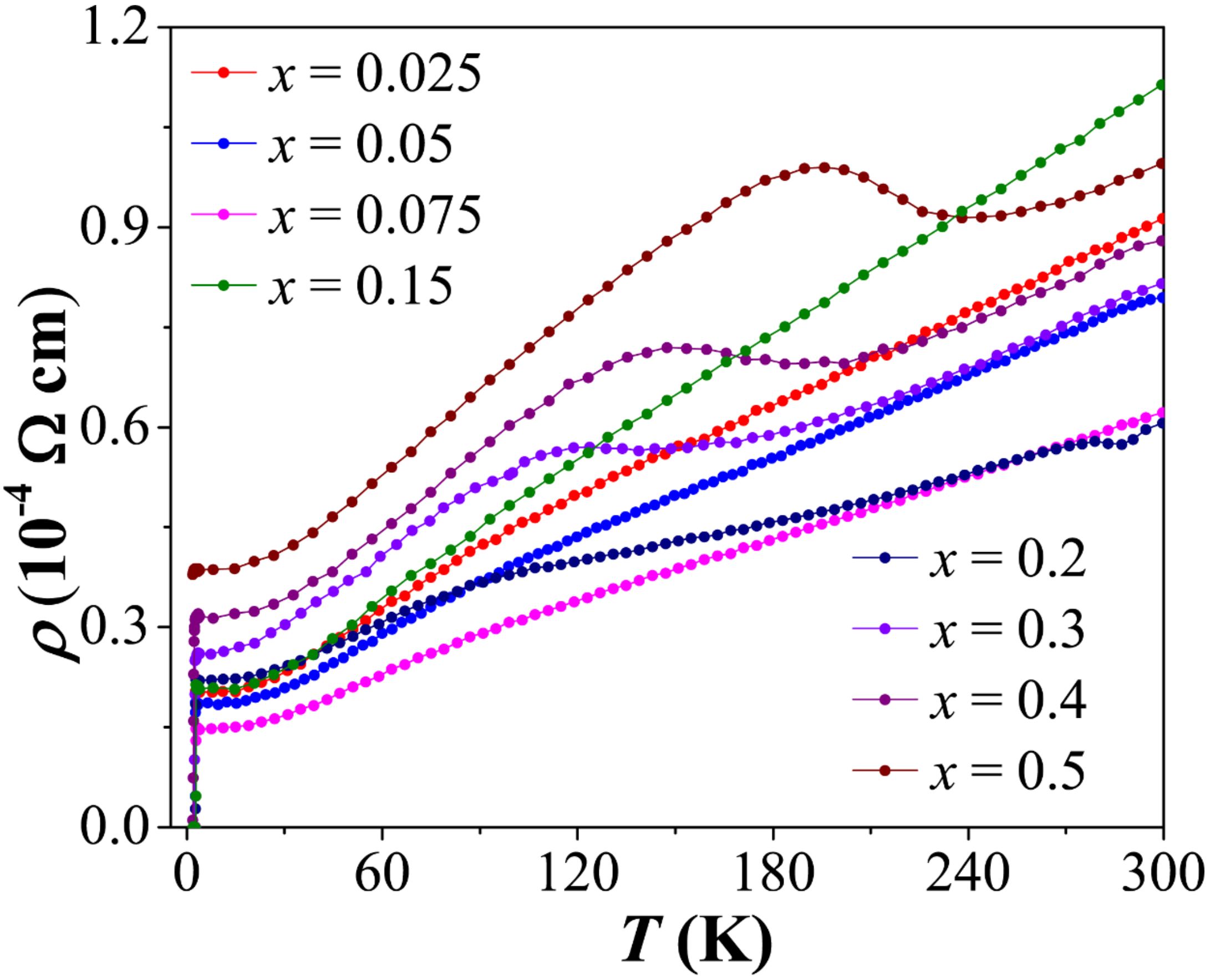

**Fig. S5.** The temperature dependent real resistivity $\rho$ (T) for polycrystalline $CuIr_2Te_{4-x}S_x$ (0.1 $\leq x \leq$ 0.5).

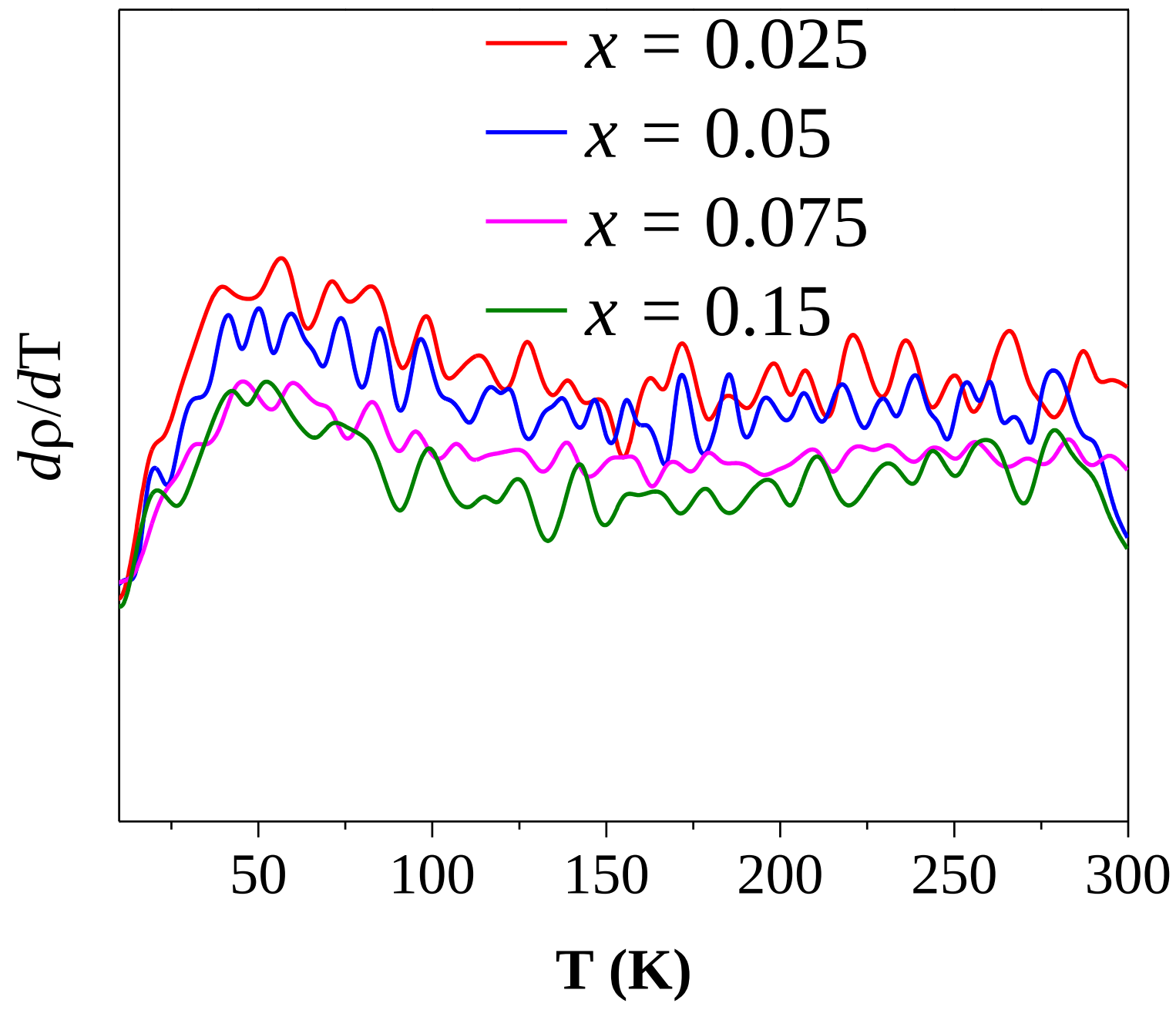

**Fig. S6.** The derivative of the resistivity data ***d*ρ/*d*T** of the low doping region samples ($x$ = 0.025, 0.05, 0.075, 0.15).

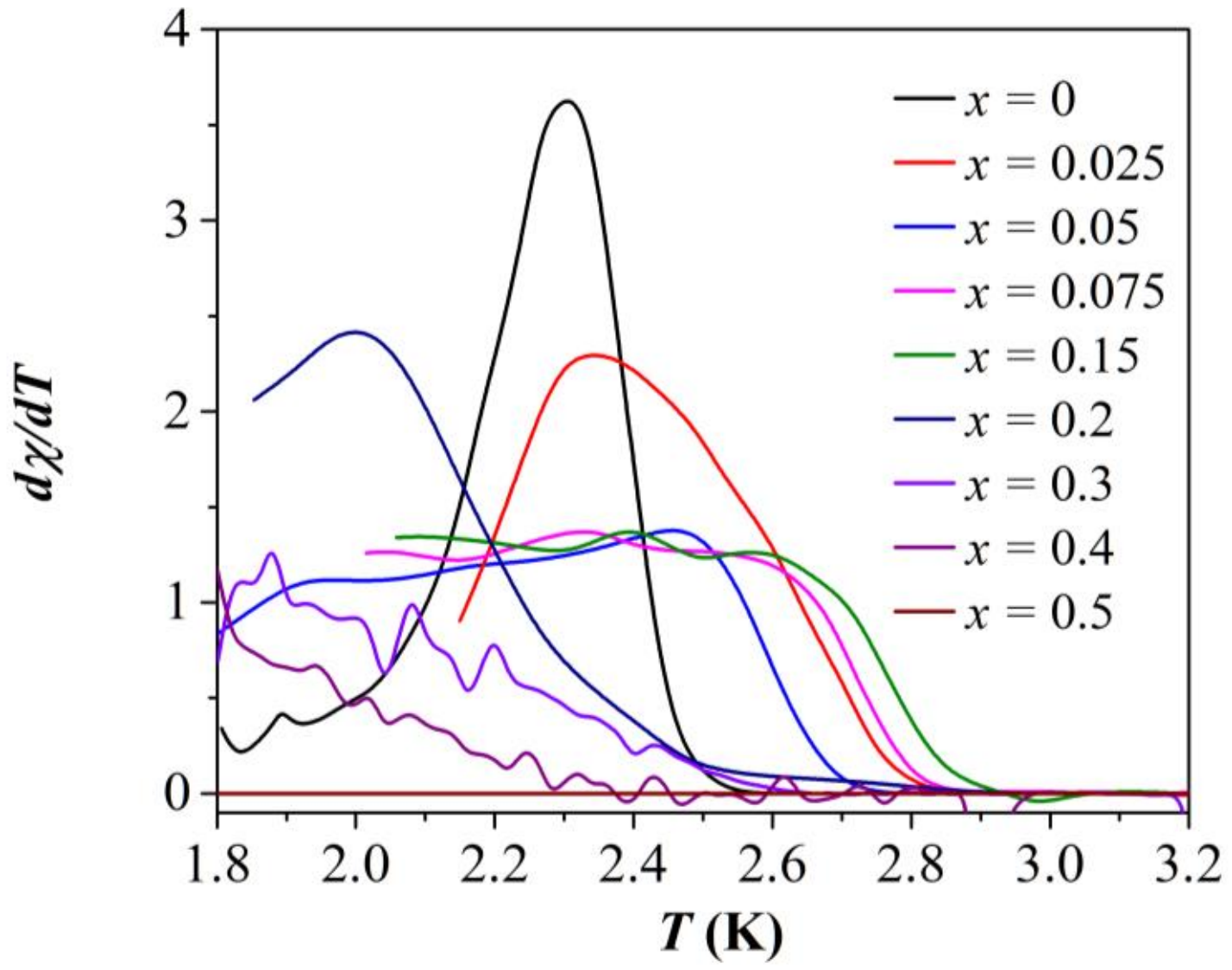

**Fig. S7.** dχ/dT vs T for the layer phase $CuIr_2Te_{4-x}S_x$ ($0.1 \le x \le 0.5$) samples.

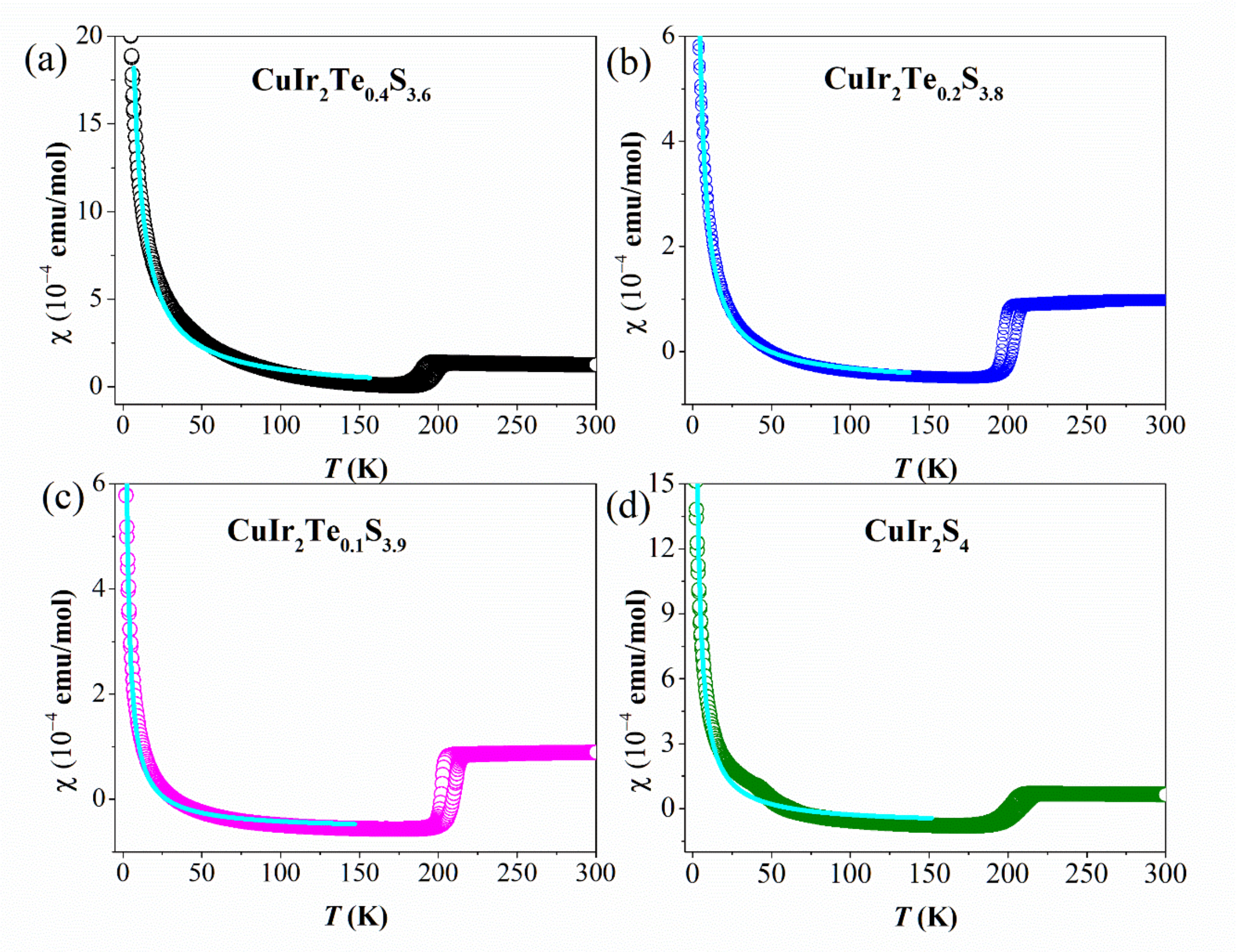

**Fig. S8.** The temperature dependence of magnetic susceptibility for (a) $CuIr_2Te_{0.4}S_{3.6}$, $CuIr_2Te_{0.2}S_{3.8}$, (c) $CuIr_2Te_{0.1}S_{3.9}$ and (d) $CuIr_2S_4$. The magnetic susceptibility between 4 K and $T_{MI}$ are fitted by $T_{MI}$: $\chi=\chi_0+\frac{C}{T}$ (solid line).